\documentclass[11pt,tightenlines,eqsecnum,floats,aps,amsmath,amssymb,nofootinbib,prd,showpacs]{revtex4}

\usepackage{setspace}
\usepackage{amsmath,amssymb,amsfonts,amsthm}
\usepackage{graphicx}
\usepackage{enumerate}
\usepackage{mathrsfs}
\usepackage{indentfirst}
\usepackage{bm}

\begin{document}

\title{The time-dependent quantum harmonic oscillator revisited:\\ Applications to quantum field theory}

\author{Daniel \surname{G\'omez Vergel}}
\email[]{dgvergel@iem.cfmac.csic.es} \affiliation{Instituto de
Estructura de la Materia, CSIC, Serrano 121, 28006 Madrid, Spain}
\author{Eduardo J. \surname{S. Villase\~nor}}
\email[]{ejsanche@math.uc3m.es} \affiliation{Grupo de Modelizaci\'on
y Simulaci\'on Num\'erica, Universidad Carlos III de Madrid, Avda.
de la Universidad 30, 28911 Legan\'es, Spain} \affiliation{Instituto
de Estructura de la Materia, CSIC, Serrano 123, 28006 Madrid, Spain}

\date{February 27, 2009}

\begin{abstract}
\indent In this article, we formulate the study of the unitary time
evolution of systems consisting of an infinite number of uncoupled
time-dependent harmonic oscillators in mathematically rigorous
terms. We base this analysis on the theory of a single
one-dimensional time-dependent oscillator, for which we first
summarize some basic results concerning the unitary implementability
of the dynamics. This is done by employing techniques different from
those used so far to derive the Feynman propagator. In particular,
we calculate the transition amplitudes for the usual harmonic
oscillator eigenstates and define suitable semiclassical states for
some physically relevant models. We then explore the possible
extension of this study to infinite dimensional dynamical systems.
Specifically, we construct Schr\"{o}dinger functional
representations in terms of appropriate probability spaces, analyze
the unitarity of the time evolution, and probe the existence of
semiclassical states for a wide range of physical systems,
particularly, the well-known Minkowskian free scalar fields and
Gowdy cosmological models.
\\
\linebreak \noindent Keywords: Time dependent harmonic oscillator,
quantum field theory, Schr\"{o}dinger representation
\end{abstract}

\pacs{03.65.-w, 03.70.+k, 04.62.+v}

\newtheorem{thr}{Theorem}
\newtheorem{lm}{Lemma}
\newtheorem{rem}{Remark}

\maketitle

%-------------------------------------------------
%                    SECTION I
%-------------------------------------------------

\section{Introduction}

\indent The quantum time-dependent harmonic oscillator (TDHO) has
received a lot of attention due to its usefulness to describe the
dynamics of many physical systems. This is the case, for example, of
radiation fields propagating outside time-dependent laser sources or
in spatial regions filled with time-dependent dielectric constant
matter \cite{Glauber1}. The behavior of ions in Paul traps
\cite{Paul,Brown,CiracGaray} can also be described by
one-dimensional harmonic oscillators with time-dependent
frequencies. The mathematical aspects of the quantum TDHO and its
applications to more general theoretical models have been profusely
analyzed in the literature \cite{Hagedorn,Gadella,LiuLi,Albeverio}.
In particular, they have been studied in the context of the search
of exact invariants for nonstationary quantum systems. This method
was introduced for the first time by Lewis and Riesenfeld
\cite{Lewis,LewisRiesenfeld} and proved to be especially useful to
generate exact solutions to the Schr\"{o}dinger equation and also to
probe the existence and properties of semiclassical states for these
systems.
\\
\indent In the context of infinite dimensional dynamical systems,
physical theoretical models with infinitely many time-dependent
oscillators naturally appear in quantum field theory in curved
spacetimes \cite{Birrell,Wald} and also in the reduced phase space
description of some midisuperspace models in general relativity
\cite{Schmidt,Mena,Pierri,Barbero2}. It is clear that, in the quest
for a suitable quantization for these systems, the understanding of
the special features of the single TDHO is particularly advisable.
This fact motivates us to summarize in a rigorous and self-contained
way the theory of the quantum TDHO in the first part of the paper by
(implicitly) making use of the theory of invariants. Although some
of the basic results such as the formula of the Feynman propagator
are certainly well-known, they will be recovered by using novel
techniques, and contrasted with the expressions obtained in the
existing literature. This procedure will facilitate the
generalization to field theories in the second part of the paper,
where we apply the ideas exposed for the one-dimensional oscillator
in order to obtain the propagator for infinite-dimensional systems
and discuss applications both to quantum field theory and quantum
gravity. The structure of the article is the following.
\\
\indent In \emph{section \ref{TDHOproperties}}, we analyze some
relevant properties of the classical TDHO equation, in particular,
its connection with the so-called Ermakov-Pinney (EP) equation
\cite{Ermakov,Pinney,ErmakovComment}, which plays an auxiliary role
in the calculation of invariants for nonquadratic Hamiltonian
systems \cite{LewisLeach}. Many of the quantities derived in this
section are used afterward to obtain a simple and closed expression
for the unitary evolution operator of the quantum TDHO.
\\
\indent In \emph{section \ref{Unitaryoperator}}, we first briefly
review the definition of the abstract Weyl $C^*$-algebra of quantum
observables for the TDHO, the uniqueness of all regular irreducible
representations of the canonical commutation relations, and the
unitary implementability of the symplectic transformations that
characterize the classical time evolution. Once a concrete
representation is fixed, we construct the unitary evolution operator
by introducing some suitable displacement and squeeze operators
\cite{MoyaCessa}. In our discussion, the use of the auxiliary EP
equation is appropriately interpreted as a natural way to solve the
Schr\"{o}dinger equation and obtain an expression of the evolution
operator valid for \emph{all} values of the time parameter. We then
show that this method is especially useful to derive the Feynman
propagator, whose calculation follows readily in this context. We
obtain an expression in agreement with those previously derived in
the existing literature, where (more complicated) path-integration
techniques are often employed \cite{Lawande1,Lawande2}. We also
analyze in this section the calculation of the transition amplitudes
for the usual harmonic oscillator eigenstates and, as a particular
case, the instability of the vacuum state as a direct consequence of
the nonautonomous nature of the system.
\\
\indent The eigenstates of the Lewis invariant \cite{Lewis} provide
a family of state vectors closed under time evolution, depending on
a particular solution to the EP equation, that generalize the
minimal wave packets of the harmonic oscillator with constant
frequency. These states, however, do not have the usual properties
of the ordinary coherent states --not even the ones associated with
the squeezed states-- and can be taken as semiclassical states just
in case they are well-behaved enough. In \emph{section
\ref{SemiclStates}}, we analyze the construction of semiclassical
states for some physically relevant systems, such as the vertically
driven pendulum and, particularly, the class of TDHO equations that
occur in the well-known Gowdy cosmological models \cite{Gowdy}, as a
previous step to generalize this construction to field theories.
\\
\indent \emph{Section \ref{FieldTheory}} is precisely devoted to the
extension of the previous study to linear dynamical systems with
infinite degrees of freedom governed by nonautonomous quantum
Hamiltonians that can be interpreted as systems of infinite
uncoupled harmonic oscillators with time-dependent frequencies. This
fact allows us to give a straightforward procedure to obtain the
unitary evolution operator, following the discussion developed for a
single oscillator. We particularize our results to the well-known
Minkowskian free scalar fields and also to the Gowdy cosmologies,
that have attracted considerable attention in recent years as
appealing frameworks to test quantum gravity theories (see
\cite{Mena,Pierri,Barbero2,Corichi,Barbero1,Barbero3} and references
therein). Making use of Schr\"{o}dinger representations, where
states act as functionals on appropriate quantum configuration
spaces, we construct the analog of the one-dimensional propagator.
We also discuss the difficulties that arise when dealing with
infinite dimensional systems -specifically, the impossibility of
unitarily implementing some symplectic transformations- and their
implications for the search of semiclassical states. We conclude the
paper with some final comments and remarks in \emph{section
\ref{Conclusions}} and \emph{appendix \ref{appendix}}.
\\
\indent Throughout the paper, we will take units such that the
Planck constant $\hbar$, the light velocity $c$, and the
characteristic mass of the system under study are equal to one. For
any $z\in\mathbb{C}\setminus(-\infty,0]\,$, $\sqrt{z}$ will denote
the unique square root of $z$ such that $\mathrm{Re}(\sqrt{z})$ is
strictly positive.

%-------------------------------------------------
%                    SECTION II
%-------------------------------------------------

\section{Properties of the TDHO equation}\label{TDHOproperties}

We will review in this section some properties of the classical
equation of motion of a single harmonic oscillator with
time-dependent frequency, from now on referred to as the TDHO
equation, given by
\begin{eqnarray}
\ddot{u}(t)+\kappa(t)u(t)=0\,,\quad t\in I=(t_-,t_+)\subseteq
\mathbb{R}\,,\label{TDHO_eq}
\end{eqnarray}
where $\kappa:I\rightarrow \mathbb{R}$ is a real-valued continuous
function and time-derivatives are denoted by dots. Given an initial
time $t_0\in I\,$, let $c_{t_0}$ and $s_{t_0}$ be the independent
solutions of (\ref{TDHO_eq}) such that
$c_{t_0}(t_0)=\dot{s}_{t_0}(t_0)=1$ and
$s_{t_0}(t_0)=\dot{c}_{t_0}(t_0)=0$. These can be written in terms
of any set of independent solutions to (\ref{TDHO_eq}), say $u_1$
and $u_2$, as
 \begin{eqnarray}\label{c&srelations}
c_{t_0}(t)
=\frac{\dot{u}_2(t_0)u_1(t)-\dot{u}_1(t_0)u_2(t)}{W(u_1,u_2)}\,,\quad
s_{t_0}(t) =\frac{u_1(t_0)u_2(t)-u_2(t_0)u_1(t)}{W(u_1,u_2)}\,,
\end{eqnarray}
where $(t_0,t)\in I\times I$ and
$W(u_1,u_2):=u_1\dot{u}_2-\dot{u}_1u_2$ denotes the
(time-independent) Wronskian of $u_1$ and $u_2$. In what follows, we
will use the notation $c(t,t_0):=c_{t_0}(t)$,
$\dot{c}(t,t_0):=\dot{c}_{t_0}(t)$, $s(t,t_0):=s_{t_0}(t)$, and
$\dot{s}(t,t_0):=\dot{s}_{t_0}(t)$. Note that the $s$ function
belongs to the class $C^2(I\times I)\,$, whereas $c(\cdot,t_0)\in
C^2(I)$ and $c(t,\cdot)\in C^1(I)\,$. As a concrete example, for the
time-independent harmonic oscillator (TIHO) with constant frequency
$\kappa(t)=\kappa_0\in \mathbb{R}$, we simply get ($\omega>0$)
\begin{eqnarray}\label{c&sMinkowski}
\kappa_0&=&\omega^2\,, \phantom{aaa}
c(t,t_0)=\cos((t-t_0)\omega)\,,\quad \phantom{a}
s(t,t_0)=\omega^{-1}\sin((t-t_0)\omega)\,;\\
\kappa_0&=&0\,, \phantom{aaaa}
c(t,t_0)=1\,,\quad\phantom{aaaaaaaaaaa,}
s(t,t_0)=t-t_0\,;\label{c&sfree}\\
\kappa_0&=&-\omega^2\,,  \phantom{a,}
c(t,t_0)=\cosh((t-t_0)\omega)\,,\quad
s(t,t_0)=\omega^{-1}\sinh((t-t_0)\omega)\,.\label{c&stachy}
\end{eqnarray}
In fact, as well known from Sturm's theory, the $c$ and $s$
functions corresponding to arbitrary frequencies share several
properties with the usual cosine and sine functions. Firstly, their
Wronskian is normalized to unit, $W(c,s)=1$. Hence, if one of them
vanishes for some time $t=t_*$, then the other is automatically
different from zero at that instant. In view of this condition and
Eq. (\ref{c&srelations}), their time-derivatives satisfy
\begin{equation}\label{derivat}
\dot{s}(t,t_0)=c(t_0,t)\,,\quad
\dot{c}(t,t_0)=\frac{c(t,t_0)c(t_0,t)-1}{s(t,t_0)}\,,
\end{equation}
where the last equation must be understood as a limit for those
values of the time parameter $t_*$ such that $s(t_*,t_0)=0$. The odd
character of the sine function translates into the condition
$s(t_0,t)=-s(t,t_0)$. Finally, the well-known formula for the sine
of a sum of angles can be generalized to
\begin{equation}\label{sum}
s(t_2,t_1)=c(t_1,t_0)s(t_2,t_0)-c(t_2,t_0)s(t_1,t_0)\,.
\end{equation}
\indent It is well known that solutions to the TDHO equation
(\ref{TDHO_eq}) are related to certain non-linear differential
equations. Here, we will restrict our attention to the so-called
Ermakov-Pinney (EP) equation (see \cite{Ermakov,Pinney}; the
interested reader is strongly advised to consult the historical
account of \cite{ErmakovComment} and references therein). Let
$$
A= \left(
  \begin{array}{cc}
    a_{11} & a_{12} \\
    a_{12} & a_{22} \\
  \end{array}
\right)
$$
be a positive definite quadratic form with $\det A=1$. Then, the
(never vanishing) function $\rho:I\rightarrow (0,+\infty)$ defined
as
\begin{eqnarray}
\rho(t):=\sqrt{a_{11}c^2(t,t_0) +a_{22}s^2(t,t_0)+2a_{12}
s(t,t_0)c(t,t_0)}\label{rho_sc}
\end{eqnarray}
satisfies the EP equation
\begin{eqnarray}
\ddot{\rho}(t)+\kappa(t)\rho(t)=\frac{1}{\rho^3(t)}\label{EP_eq}\,,\quad
t\in I\,.
\end{eqnarray}
According to Eq. (\ref{c&srelations}), the most general analytic
solution to Eq. (\ref{EP_eq}) can be written as
\cite{LewisErmakov,Lutzky}
\begin{equation}\label{rhoABC}
\rho(t)=\sqrt{b_{11}u_1^2(t)+b_{22}u_2^2(t)+2b_{12}u_1(t)u_2(t)}\,,
\end{equation}
where, as a consequence of (\ref{rho_sc}) and (\ref{EP_eq}), the
coefficients  $b_{11}$, $b_{12}$, $b_{22}\in\mathbb{R}$ satisfy
$W^2(u_1,u_2)=(b_{11}b_{22}-b_{12}^2)^{-1}>0$. Conversely, given
\emph{any} solution  to the EP equation it is possible to find the
general solution to the TDHO equation. Indeed, it is straightforward
to prove the following theorem.

\begin{thr}\label{ThrEP}

Let $\rho$ be any solution to the EP equation (\ref{EP_eq}); then,
the $c$ and $s$ solutions to (\ref{TDHO_eq}) are given by
\begin{eqnarray}
c(t,t_0)&=&\frac{\rho(t)}{\rho(t_0)}\cos\left(\int_{t_0}^t\frac{\mathrm{d}\tau}{\rho^2(\tau)}\right)
-\rho(t)\dot{\rho}(t_0)\sin\left(\int_{t_0}^t\frac{\mathrm{d}\tau}{\rho^2(\tau)}\right)\,,\label{c_rho}\\
s(t,t_0)&=&\rho(t)\rho(t_0)\sin\left(\int_{t_0}^t\frac{\mathrm{d}\tau}{\rho^2(\tau)}\right)\,,\label{s_rho}\quad
(t,t_0)\in I\times I\,.
\end{eqnarray}
\end{thr}
\begin{rem} \emph{By using Eq. (\ref{c_rho}) and
(\ref{s_rho}), it is possible to find other $\rho$-independent
objects. For example, the combination
$$
\frac{\rho(t_0)}{\rho(t)}\cos\left(\int_{t_0}^t\frac{\mathrm{d}\tau}{\rho^2(\tau)}\right)
+\rho(t_0)\dot{\rho}(t)\sin\left(\int_{t_0}^t\frac{\mathrm{d}\tau}{\rho^2(\tau)}\right)=c(t_0,t)=\dot{s}(t,t_0)
$$
and the zeros of $s(t,t_0)$, characterized by
$$
\int_{t_0}^t\frac{\mathrm{d}\tau}{\rho^2(\tau)}\equiv 0 \,\,\,
(\textrm{mod}\, \pi)\,,
$$
are independent of the particular solution $\rho$ to the EP
equation. These results will be profusely applied along the
article.}
\end{rem}

%-------------------------------------------------
%                  SECTION III
%-------------------------------------------------

\section{Unitary quantum time evolution}\label{Unitaryoperator}

\subsection{General framework}

The \emph{canonical} phase space description of the classical system
under consideration consists of a nonautonomous Hamiltonian system
$(I\times\Gamma,\mathrm{d}t,\bm{\omega},H(t))$. Here,
$\Gamma:=\mathbb{R}^2$ denotes the space of Cauchy data $(q,p)$
endowed with the usual symplectic structure
$\bm{\omega}((q_1,p_1),(q_2,p_2)):=p_1q_2-p_2q_1$,
$\forall\,(q_1,p_1), (q_2,p_2)\in\Gamma$. The triplet
$(I\times\Gamma,\mathrm{d}t,\bm{\omega})$ then has the mathematical
structure of a cosymplectic vector space (see \cite{Marsden} for
more details). The time-dependent Hamiltonian
$H(t):\Gamma\rightarrow\mathbb{R}$, $t\in I$, is given by
\begin{equation}\label{Hclass}
H(t,q,p):=\frac{1}{2} \Big(p^2+\kappa(t)q^2\Big)\,.
\end{equation}
The solution to the corresponding Hamilton equations with initial
Cauchy data $(q,p)$ at time $t_0$ can be written down as
\begin{eqnarray}
\left(\begin{array}{c}
                 q_H(t,t_0) \\
                 p_H(t,t_0) \\
               \end{array}
             \right)=\mathcal{T}_{(t,t_0)}\cdot\left(
               \begin{array}{c}
                 q \\
                 p \\
               \end{array}
             \right),\quad \mathcal{T}_{(t,t_0)}:=\left(
        \begin{array}{cc}
          c(t,t_0) & s(t,t_0) \\
       \dot{c}(t,t_0)& \dot{s}(t,t_0) \\
        \end{array}
      \right). \label{Hamilton_class}
\end{eqnarray}
Note that the properties stated in \emph{section
\ref{TDHOproperties}} about the $c$ and $s$ solutions to the TDHO
equation (\ref{TDHO_eq}) guarantee that $\mathcal{T}_{(t,t_0)}\in
SL(2,\mathbb{R})=SP(1,\mathbb{R})$ for all $(t,t_0)\in I\times I\,$,
i.e., the classical time evolution is implemented by symplectic
transformations.
\\
\indent We now formulate the quantum theory of the TDHO by defining
an appropriate abstract $C^*$-algebra of quantum observables
\cite{Neumann}. This algebraic approach, although being more
complicated than the traditional canonical quantization for this
system, will facilitate the study of field theories in subsequent
sections. We first realize that, as a consequence of the linearity
of $\Gamma$, the set of elementary classical observables can be
identified with the $\mathbb{R}$-vector space generated by linear
functionals on $\Gamma$. Every pair $\lambda:=(-b,a)\in\Gamma$ has
an associated functional $F_\lambda:\Gamma\rightarrow\mathbb{R}$
such that, for all $X=(q,p)\in\Gamma$,
$F_\lambda(X):=\bm{\omega}(\lambda,X)=a q+b p\,$. In order to
quantize the system, we introduce the abstract Weyl $C^*$-algebra of
quantum observables on $\Gamma$, $\mathscr{W}(\Gamma)$, generated by
the unitary Weyl operators $W(\lambda)=\exp(iF_\lambda)$,
$\lambda\in\Gamma$, satisfying
\begin{equation}\label{Weyl}
W(\lambda_1)^{*}=W(-\lambda_1)\,,\quad
W(\lambda_1)W(\lambda_2)=\exp\left(i\bm{\omega}(\lambda_1,\lambda_2)/2\right)
W(\lambda_1+\lambda_2)\,,\,\,\,\forall\,\lambda_1,\,\lambda_2\in\Gamma\,.
\end{equation}
According to von Neumann's uniqueness theorem
\cite{Neumann,Prugovecki}, all regular irreducible representations
$\pi:\mathscr{W}(\Gamma)\rightarrow\mathscr{B}(\mathscr{H})$ of the
Weyl $C^*$-algebra into separable Hilbert spaces
$(\mathscr{H},\langle\cdot\,|\,\cdot\rangle)$ are unitarily
equivalent. Here, $\mathscr{B}(\mathscr{H})$ denotes the collection
of bounded linear operators on $\mathscr{H}$. A $*$-homomorphism
$\pi:\mathscr{W}(\Gamma)\rightarrow\mathscr{B}(\mathscr{H})$ is said
to be a regular irreducible representation if it has $\{0\}$ and
$\mathscr{H}$ as the only closed $\pi$-invariant subspaces, and
$\pi(W(0,a))$ and $\pi(W(-b,0))$ are strongly continuous in the $a$
and $b$ parameters, respectively. A well-known solution is given by
the Schr\"{o}dinger representation
$\pi_{s}:\mathscr{W}(\Gamma)\rightarrow\mathscr{B}(L^2(\mathbb{R},\mathrm{d}q))$
into the Hilbert space $L^2(\mathbb{R},\mathrm{d}q)$ where, for all
pure states  $\psi\in L^2(\mathbb{R},\mathrm{d}q)$,
$$(\pi_s(W(\lambda))\cdot\psi)(q):=\exp\left(-iab/2\right)\exp(ia q)\psi(q+b)\,,\quad\lambda=(-b,a)\in\Gamma\,.$$
Thanks to the regularity condition, the usual Heisenberg algebra can
be recovered in a definite sense from the Weyl $C^{*}$-algebra. The
strong continuity of $\pi_s(W(0,a))$ and $\pi_s(W(-b,0))$ in the
real variables $a$ and $b$ ensures, by virtue of Stone's theorem,
the existence of (unbounded) self-adjoint generators $Q$ and $P$
with dense domains in $L^2(\mathbb{R},\mathrm{d}q)$. In particular,
the Schwartz space $\mathscr{S}(\mathbb{R})$ of smooth rapidly
decreasing functions in $\mathbb{R}$ is a common invariant dense
domain of essential self-adjointness for $Q$ and $P$, where the
usual Heisenberg algebra is satisfied. For all
$\psi\in\mathscr{S}(\mathbb{R})$, we have $(Q\psi)(q)=q\psi(q)$ and
$(P\psi)(q)=-i\psi^\prime(q)$, where $\psi^\prime$ denotes the
derivative of $\psi$.
\\
\indent Another possibility is to represent the canonical
commutation relations (CCR) in the space
$L^{2}(\mathbb{R},\mathrm{d}\mu_{\alpha})$ where, given some
$\alpha\in\mathbb{C}\setminus\{0\}$, $\mu_{\alpha}$ denotes the
Gaussian probability measure
$$
\mathrm{d}\mu_{\alpha}=\frac{1}{\sqrt{2\pi}|\alpha|}\exp\left(-\frac{q^2}{2|\alpha|^2}\right)\,\mathrm{d}q\,.
$$
To each $\alpha$ there corresponds a family of unitary
transformations
$V_\alpha(\beta):L^2(\mathbb{R},\mathrm{d}q)\rightarrow
L^{2}(\mathbb{R},\mathrm{d}\mu_{\alpha})$ connecting the standard
Hilbert space with the new one in the form
\begin{eqnarray}\label{Valphabeta}
\Psi(q)=\big(V_\alpha(\beta)\psi\big)(q)=\big(\sqrt{2\pi}|\alpha|\big)^{1/2}\exp\big(-i\bar{\beta}q^2/(2\bar{\alpha})\big)\psi(q)\,,
\end{eqnarray}
where the complex numbers $\beta$ must satisfy
$\alpha\bar{\beta}-\beta\bar{\alpha}=i\,$. Note that the unitary
transformations $V_\alpha(\beta)$ map the `vacuum' state
$$\psi_0(q)=\big(\sqrt{2\pi}|\alpha|\big)^{-1/2}\exp\big(i\bar{\beta}q^2/(2\bar{\alpha})\big)\in
L^{2}(\mathbb{R},\mathrm{d}q)$$ into the unit function
$\Psi_0(q)=(V_\alpha(\beta)\psi_0)(q)=1\in
L^{2}(\mathbb{R},\mathrm{d}\mu_\alpha)\,$. In these cases, the
position and momentum operators act on state vectors as
$$(Q\Psi)(q)=q\Psi(q)\quad \mathrm{and}\quad (P\Psi)(q)=-i\Psi'(q)+\frac{\bar{\beta}}{\bar{\alpha}}q\Psi(q)\,,$$
where, with the aim of simplifying the notation, $Q$ and $P$
respectively denote the Schr\"{o}dinger transformed operators
$V_\alpha(\beta)QV_\alpha(\beta)^{-1}$ and
$V_\alpha(\beta)PV_\alpha(\beta)^{-1}$ with common dense domain
$V_\alpha(\beta)\mathscr{S}(\mathbb{R})\subset
L^{2}\big(\mathbb{R},\mathrm{d}\mu_\alpha\big)\,$.
\\
\indent Any regular irreducible representation
$\pi:\mathscr{W}(\Gamma)\rightarrow\mathscr{B}(\mathscr{H})$ is
stable under time evolution, i.e., there exists a (biparametric)
family of unitary operators
$U(t,t_0):\mathscr{H}\rightarrow\mathscr{H}$, the so-called quantum
time evolution operator, such that
\begin{equation}\label{Urelations}
U^{-1}(t,t_0)\,\pi(W(\lambda))\,U(t,t_0)=\pi(W(\mathcal{T}_{(t,t_0)}\lambda))
\,,\,\,\,\,\forall\,W(\lambda)\in\mathscr{W}(\Gamma)\,,\,\,\,\forall\,(t,t_0)\in
I\times I\,,
\end{equation}
with $\mathcal{T}_{(t,t_0)}$ defined in Eq. (\ref{Hamilton_class}).
These relations determine $U(t,t_0)$ univocally up to phase. It is
important to notice at this point that, if the classical evolution
has singularities at the boundary of the interval $I$, they also
occur for the quantum dynamics, i.e., there is no resolution of
classical singularities. On the other hand, we will check in
\emph{section \ref{FieldTheory}} that the unitary implementability
of symplectic transformations like those corresponding to the
classical time evolution is not directly guaranteed for
infinite-dimensional systems. The Heisenberg equations for $Q$ and
$P$ can be solved just by the same expressions involved in the
classical solutions (\ref{Hamilton_class}), i.e.,
\begin{eqnarray}\label{QHPH}
\left(
  \begin{array}{c}
    Q_H(t,t_0) \\
    P_H(t,t_0)
  \end{array}
\right)&:=&U^{-1}(t,t_0)\left(
  \begin{array}{c}
    Q \\
    P
  \end{array}
\right) U(t,t_0)=\left(
        \begin{array}{cc}
          c(t,t_0) & s(t,t_0) \\
          \dot{c}(t,t_0)& \dot{s}(t,t_0) \\
        \end{array}
      \right) \left(
  \begin{array}{c}
    Q \\
    P
  \end{array}
\right).
\end{eqnarray}
With more generality, given any well-behaved (analytic) classical
observable $F:\Gamma\rightarrow\mathbb{R}$ for the TDHO, the time
evolution of its quantum counterpart
$F_{H}(t,t_0):=U^{-1}(t,t_0)F(Q,P)U(t,t_0)$ in the Heisenberg
picture is simply given by
\begin{equation}
F_H(t,t_0)=F(Q_H(t,t_0),P_H(t,t_0))
=F\big(c(t,t_0)Q+s(t,t_0)P,\dot{c}(t,t_0)Q+
\dot{s}(t,t_0)P\big)\,.\label{Heisen}
\end{equation}
Hence, the matrix elements $\langle
\Psi_2\,|\,U^{-1}(t_2,t_1)F(Q,P)U(t_2,t_1) \Psi_1 \rangle$,
$\Psi_1,\Psi_2\in\mathscr{H}$, can be computed without the explicit
knowledge of the unitary evolution operator. This is also the case
for the probability transitions
$\mathrm{Prob}(\Psi_2,t_2\,|\,\Psi_1,t_1)=|\langle\Psi_2\,|\,U(t_2,t_1)\Psi_1
\rangle|^2$, as will be discussed in detail in \textit{subsection
\ref{TransInstab}}. The commutators of time-evolved observables can
be also calculated without the concrete expression of $U(t_2,t_1)$.
For instance, from Eq. (\ref{Heisen}) we easily obtain
$$\big[Q_{H}(t_1,t_0),Q_{H}(t_2,t_0)\big]=is(t_1,t_2)\mathbf{1}\,,$$
where we have used the relation (\ref{sum}) stated in \emph{section
\ref{TDHOproperties}}. As expected, the commutator given above is
proportional to the identity operator and independent of the choice
of the initial time $t_0\,$. Note, in contrast with the transition
probabilities, that the calculation of transition amplitudes of the
type $\langle \Psi_2\,|\, U(t_2,t_1)\Psi_1\rangle$ does require the
explicit knowledge of (the phase of) the evolution operator. This is
also the case for the (strong) derivatives of both $U(\cdot,t_0)$
and $U(t,\cdot)\,$.
\\
\indent The dynamics of the quantum TDHO is governed by an
(unbounded) nonautonomous Hamiltonian operator
$H(t):\mathscr{H}\rightarrow\mathscr{H}$, $t\in I$, satisfying
\begin{equation}
\dot{U}(t,t_0)=-iH(t)U(t,t_0)\,.\label{Uequa}
\end{equation}
Given the quadratic nature of the classical Hamiltonian
(\ref{Hclass}), $H(t)$ must coincide with the operator directly
promoted from the classical function modulo a $t$-dependent real
term proportional to the identity $\mathbf{1}$ that encodes the
election of $U(t,t_0)$ satisfying Eq. (\ref{Urelations}). For a
concrete representation of the CCR, we will simply take
\begin{equation}\label{quantHamiltonian}
H(t):=\frac{1}{2}\left(P^{2}+\kappa(t)Q^{2}\right).
\end{equation}
This choice fixes $U(t,t_0)$ uniquely. The Hamiltonian
(\ref{quantHamiltonian}) is a self-adjoint operator with dense
domain $\mathscr{D}_{H(t)}$ --equal to $C_0^\infty(\mathbb{R})$ in
the standard Schr\"{o}dinger representation-- for each value of the
time parameter $t\,$. We will prove the following theorem in the
next subsections.

\begin{thr}\label{ThrPropag} The action of the unitary TDHO evolution operator $U(t,t_0)$ corresponding
to the Hamiltonian (\ref{quantHamiltonian}) on any state vector
$\psi\in\mathscr{S}(\mathbb{R})\subset
L^{2}(\mathbb{R},\mathrm{d}q)$ in the traditional Schr\"{o}dinger
representation is given by
$$
\big(U(t,t_0)\psi\big)(q)=\int_{\mathbb{R}}
K(q,t;q_0,t_0)\psi(q_0)\,\mathrm{d}q_0\,,
$$
where the propagator $K(q,t;q_0,t_0)$ depends on the times $t_0$ and
$t$ through the classical TDHO solutions $c$ and $s\,$. Explicitly,
\begin{eqnarray}
K(q,t;q_0,t_0)&=&\frac{1}{\sqrt{2\pi
i}}\,s^{-1/2}(t,t_0)\exp\left(\frac{i}{2
s(t,t_0)}\Big(c(t_0,t)q^2+c(t,t_0)q_0^2-2qq_0\Big)\right),\label{K_qq0a}
\end{eqnarray}
wherever $s(t,t_0)\neq 0\,$, and
\begin{eqnarray}
K(q,t;q_0,t_0) =
c^{-1/2}(t,t_0)\exp\left(i\frac{\dot{c}(t,t_0)}{2c(t,t_0)}\right)\label{K_qq0b}
\delta(q_0-q/c(t,t_0))
\end{eqnarray}
if $s(t,t_0)=0\,$.
\end{thr}

\begin{rem}
\emph{Given a solution $u(t)$ to the TDHO equation (\ref{TDHO_eq})
which is positive in some interval $(t_0,t_0+\varepsilon)\subset I$,
$\varepsilon>0$, we define
$$
u^{\epsilon}(t,t_0):=\exp\big(i\epsilon\pi
\,\mathfrak{m}(u;t,t_0)\big)|u(t)|^{\epsilon}\,,\,\,\,\epsilon\in\mathbb{R}\,,\,\,\,t\in
I\,,
$$
where $\mathfrak{m}(u;t,t_0)\in \mathbb{Z}$ is the index function of
$u$, with  $\mathfrak{m}(u;t_0,t_0)=0\,$, defined in such a way that
$\mathfrak{m}(u;t_2,t_0)-\mathfrak{m}(u;t_1,t_0)$, $t_1<t_2$, gives
the number of zeros of $u(\cdot,t_0)$ in the interval $(t_1,t_2]\,$.
Finally, $\delta(q)$ denotes the Dirac delta distribution.}
\end{rem}
\begin{rem} \emph{Let $\vartheta:I\rightarrow \mathbb{R}$ be a real-valued continuous function and consider the Hamiltonian
$$H_1(t):=H(t)+\vartheta(t)\mathbf{1}\,$$
defined in terms of (\ref{quantHamiltonian}). The unitary evolution
$U_1(t,t_0)$ associated with $H_1(t)$ satisfying Eq.
(\ref{Urelations}) gives rise to the propagator
\begin{eqnarray}
K_1(q,t;q_0,t_0)=K(q,t:q_0,t_0)\exp\left(-i\int_{t_0}^t\vartheta(\tau)\mathrm{d}\tau\right).\label{K1}
\end{eqnarray}
Note that
$U^{-1}_1(t,t_0)\,\mathcal{O}\,U_1(t,t_0)=U^{-1}(t,t_0)\,\mathcal{O}\,U(t,t_0)$
for any quantum observable $\mathcal{O}$.}
\end{rem}
\begin{rem}
\emph{In the $L^{2}(\mathbb{R},\mathrm{d}\mu_\alpha)$-representation
defined by the unitary transformation $V_\alpha(\beta)$ (see Eq.
(\ref{Valphabeta})), the evolution is given by
$$
\big(U(t,t_0)\Psi\big)(q)=\int_{\mathbb{R}}
K_{\alpha\beta}(q,t;q_0,t_0)\Psi(q_0)\,\mathrm{d}\mu_\alpha(q_0)\,,
$$
where
\begin{eqnarray}\label{Kalphabeta}
K_{\alpha\beta}(q,t;q_0,t_0):=\sqrt{2\pi}|\alpha|\exp\left(
\frac{i\beta}{2\alpha}q_0^2-\frac{i\bar{\beta}}{2\bar{\alpha}}q^2\right)
K(q,t;q_0,t_0)\,.
\end{eqnarray}}
\end{rem}

\subsection{Constructing the evolution operator}\label{ConstructingU}

\indent  In order to calculate the unitary evolution operator
$U(t,t_0)$ we will perform a gene\-ralization of the method
developed in \cite{MoyaCessa} that will clarify the appearance of
the auxiliary Ermakov-Pinney solution (\ref{rhoABC}) in this
context, and will allow us also to warn the reader about other
problematic choices that have appeared before in the related
literature. We first introduce on $\mathscr{H}$ the (one-parameter
family of) unitary operators
$$D(x):=\exp\left(-\frac{i}{2} x Q^{2}\right),\,\,\,x\in \mathbb{R}\,,$$
generating a displacement of the momentum operator, $D(x) P
D^{-1}(x)=P+xQ$ (the position operator being unaffected by them),
and define the unitary squeeze operators
$${S}(y):=\exp\left(\frac{i}{2}y\big({Q}{P}+{P}{Q}\big)\right),\,\,\,y\in\mathbb{R}\,,$$
scaling both the position and momentum operator as
$S(y){Q}S^{-1}(y)=e^{y}{Q}$ and $S(y){P}{S}^{-1}(y)=e^{-y} P$,
respectively. Let $\Psi(t)\in\mathscr{D}_{H(t)}$, $t\in I$, be a
solution to the Schr\"{o}dinger equation, i.e.,
$i\dot{\Psi}(t)=H(t)\Psi(t)$, and let $x,y\in C^1(I)\,$. We now
introduce the unitary operators
$$T(t)=T(t;x,y):=S(y(t))D(x(t))\,,$$
where the functions $x$ and $y$ remain arbitrary at this stage. Let
us consider the time evolution for the transformed state vector
$$
\Phi(t)=\Phi(t;x,y):=T(t;x,y)\Psi(t)\,,
$$
given by
\begin{eqnarray*}
i\dot{\Phi}(t)&=&\Big(T(t)H(t)T^{-1}(t)-iT(t)\dot{T}(t)\Big)\Phi(t)\\
&=&\frac{1}{2}\Big(e^{-2y(t)}P^2+(x(t)-\dot{y}(t))(QP+PQ)
+e^{2y(t)}(x^2(t)+\kappa(t)+\dot{x}(t))Q^2\Big)\Phi(t)\,.
\end{eqnarray*}
We note at this point that it is possible to get a notable
simplification of the previous expression just by imposing
\begin{equation}\label{naturalconditions}
x(t)=\dot{y}(t)\quad\mathrm{and}\quad
x^2(t)+\kappa(t)+\dot{x}(t)=\exp(-4y(t))\,.
\end{equation}
The most natural way to achieve this is to choose
$$
y(t):=\log\rho(t)\quad \textrm{and, hence,}\quad
x(t)=\dot{\rho}(t)/\rho(t)\,,
$$
with $\rho$ being \emph{any} solution to the auxiliary EP equation
(\ref{EP_eq}) introduced in \emph{section \ref{TDHOproperties}}. In
this way, the state vector
$\Phi(t;\dot{\rho}/\rho,\log\rho)=:\Phi_\rho(t)$ satisfies the
differential equation
$$i\dot{\Phi}_{\rho}(t)=\frac{1}{2\rho^2(t)}\big(P^2+Q^2\big)\Phi_\rho(t)\,.$$
Solving this equation and going back to the original state vector
$\Psi(t)$, we finally obtain the unitary evolution operator for the
system. We can then enunciate the following theorem.

\begin{thr}\label{ThrU}
The time evolution operator $U(t,t_0)$ for the quantum TDHO whose
dynamics is governed by the Hamiltonian (\ref{quantHamiltonian}) is
given by a composition of unitary operators
$$ U(t,t_0)=T_\rho^{-1}(t)R_\rho(t,t_0)T_\rho(t_0)\,,$$
where
\begin{eqnarray}
R_\rho(t,t_0):=\exp\left(-\frac{i}{2}\int_{t_0}^t\frac{\mathrm{d}\tau}{\rho^2(\tau)}\big(P^2+Q^2\big)\right),\label{U1}
\end{eqnarray}
and $T_\rho(t)=S_\rho(t)D_\rho(t)$, with
\begin{eqnarray}
D_\rho(t):=\exp\left(-\frac{i}{2}\frac{\dot{\rho}(t)}{\rho(t)}Q^{2}\right)\quad
\mathrm{and}\quad
S_\rho(t):=\exp\left(\frac{i}{2}\log\rho(t)\big({Q}{P}+{P}{Q}\big)\right).\label{U2}
\end{eqnarray}
\end{thr}
\begin{rem} \emph{Note that instead of introducing $\rho$, we could have used
other choices for the $x$ and $y$ functions. In these cases,
conditions (\ref{naturalconditions}) may not hold and the
expressions of the evolution operator would differ from the one
obtained here. For instance, one can select $x(t)=\dot{u}(t)/u(t)$
and $y(t)=\log u(t)$ as in \cite{MoyaCessa}, with $u(t)$ being any
solution to the TDHO equation, but this choice is problematic
because the set $\{t\in I\,|\,u(t)=0 \}$ must be non-empty and,
hence, the resulting formula for the unitary operator is generally
not well-defined for all values of the time parameter $t$. This is
the reason why the election of the Ermakov-Pinney solution is
especially convenient in this context --recall that $\rho$ is a
positive definite function. It follows from the above argument that
the appearance of this solution is nearly unavoidable in this
context.}
\end{rem}

\indent Note that the eigenstates of the $R_\rho(t,t_0)$ operator
(\ref{U1}) are given by those of the Hamiltonian operator
corresponding to a quantum harmonic oscillator with unit frequency
$\sqrt{\kappa(t)}=1$,
\begin{equation}\label{H0}
H_0:=\frac{1}{2}\big(P^2+Q^2\big)\,.
\end{equation}
This fact will be shown to be particularly useful to calculate the
Feynman propagator. It is also important to point out that the
procedure employed in this section is implicitly based upon the
transformation of the so-called Lewis invariant \cite{Lewis}
\begin{equation}\label{Lewis}
I_\rho(t):=\frac{1}{2}\left(\frac{Q^{2}}{\rho^2(t)}+\big(\rho(t)P-\dot{\rho}(t)Q\big)^{2}\right)\,,\quad
\dot{I}_{\rho_H}=0\,,
\end{equation}
into an explicitly time-independent quantity --although in order to
obtain the unitary operator it has not been necessary to use it. In
this case, we simply have
\begin{equation}\label{TIT-1}
T_\rho(t)I_\rho(t) T_\rho^{-1}(t)=H_0\,.
\end{equation}
The Lewis invariant is often used to generate exact solutions to the
Schr\"{o}dinger equation, and turns out to be especially useful to
construct semiclassical states for these systems, as will be
discussed later.

\subsection{Propagator formula}

We finally proceed to derive the Feynman propagator for the quantum
TDHO corresponding to the Hamiltonian (\ref{quantHamiltonian}). In
the previous subsection, we have written down the evolution operator
for this system explicitly in closed form in terms of the position
and momentum operators (see \emph{theorem \ref{ThrU}}). It is given
by the product of the unitary operators (\ref{U1}) and (\ref{U2}).
We calculate now the action of these factors on test functions
$\psi\in\mathscr{S}(\mathbb{R})\subset L^2(\mathbb{R},\mathrm{d}q)$
in the standard Schr\"{o}dinger representation. First, it is
straightforward to see that
\begin{eqnarray*}
\big(T_\rho(t)\psi\big)(q)&=&\sqrt{\rho(t)}\exp\left( - \frac{i}{2}\dot{\rho}(t)\rho(t)q^2\right)\psi(\rho(t)q)=\int_\mathbb{R} K_\rho^+(q,t;q_0) \psi(q_0)\,\mathrm{d}q_0\,,\\
\big(T^{-1}_\rho(t)\psi\big)(q)&=&\frac{1}{\sqrt{\rho(t)}}\exp\left(
\frac{i}{2}\frac{\dot{\rho}(t)}{\rho(t)}q^2\right)\psi(q/\rho(t))=\int_\mathbb{R}
K_\rho^-(q,t;q_0) \psi(q_0)\,\mathrm{d}q_0\,,
\end{eqnarray*}
where we have introduced the distributions
\begin{eqnarray}
K_\rho^+(q,t;q_0)&:=&\sqrt{\rho(t)}\exp\left( - \frac{i}{2}\dot{\rho}(t)\rho(t)q^2\right)\delta(q_0-\rho(t)q)\,,\label{K-}\\
K_\rho^-(q,t;q_0)&:=&\frac{1}{\sqrt{\rho(t)}}\exp\left(
\frac{i}{2}\frac{\dot{\rho}(t)}{\rho(t)}q^2\right)\delta(q_0-q/\rho(t))\,.\label{K+}
\end{eqnarray}
The propagator for $R_\rho(t,t_0)$, satisfying
$$
\big(R_\rho(t,t_0)\psi\big)(q)=\int_\mathbb{R}
K_\rho^0(q,t;q_0,t_0)\psi(q_0)\, \mathrm{d}q_0\,,
$$
can be easily derived from the one corresponding to the TIHO with
unit frequency. As is well known \cite{Felsager,Rezende1}, the Green
function $K^0$ for the Hamiltonian (\ref{H0}) is given by the
Feynman-Soriau formulae
\begin{eqnarray*}
K^0(q,\upsilon;q_0,0)&=&\frac{1}{\sqrt{2\pi i}}\sin^{-1/2}(\upsilon,0)\exp\left(\frac{i}{2\sin\upsilon}\Big((q^2+q_0^2)\cos\upsilon-2qq_0\Big)\right)\,,\quad \upsilon\not\equiv 0 \,\,\,(\mathrm{mod}\,\pi)\,,\\
K^0(q,\upsilon
;q_0,0)&=&\cos^{-1/2}(\upsilon,0)\exp\left(-\frac{i\sin
\upsilon}{2\cos\upsilon}\right)\delta(q_0-q/\cos\upsilon)\,,\quad
\upsilon\equiv 0 \,\,\,(\mathrm{mod}\,\pi)\,,
\end{eqnarray*}
where the so-called Maslov correction factor \cite{Rezende1}, which
allows the calculation of the propagator beyond the caustics
$\{\upsilon\in \mathbb{R}\,:\,\sin(\upsilon)=0\}=\{\pi k\,:\, k\in
\mathbb{Z}\}\,$, has been conveniently absorbed into the definition
of $\sin^{1/2}(\upsilon,0)$ and $\cos^{1/2}(\upsilon,0)$ given in
the formulation of \emph{theorem \ref{ThrPropag}}. In view of Eq.
(\ref{U1}), we simply get
\begin{eqnarray}
K_\rho^0(q,t;q_0,t_0)=K^0\left(q,\int_{t_0}^t\frac{\mathrm{d}\tau}{\rho^2(\tau)}\,;q_0,0\right).\label{Kr}
\end{eqnarray}
Therefore,
$$
\big(U(t,t_0)\psi\big)(q)=\big(T^{-1}_\rho(t)
R_\rho(t,t_0)T_\rho(t_0)\psi\big)(q)=\int_\mathbb{R}
K(q,t;q_0,t_0)\psi(q_0)\, \mathrm{d}q_0\,,
$$
where
\begin{eqnarray}
K(q,t;q_0,t_0)&=&\int_{\mathbb{R}^2}
K_\rho^-(q,t;q_2)K_\rho^0(q_2,t;q_1,t_0)K_\rho^+(q_1,t_0;q_0)\,\mathrm{d}q_1\,\mathrm{d}q_2\,.\label{K}
\end{eqnarray}
By combining (\ref{K-})-(\ref{K}) with (\ref{c_rho}) and
(\ref{s_rho}), we find the formula for the propagator (\ref{K_qq0a})
enunciated in \emph{theorem \ref{ThrPropag}} expressed in terms of
the $c$ and $s$ solutions to the classical TDHO equations
(\ref{TDHO_eq}). As expected, the propagator --and hence the
evolution operator itself-- does not depend on the particular
solution $\rho$ to the EP equation (\ref{EP_eq}) chosen to factorize
$U(t,t_0)$. Taking the appropriate limits one obtains, after
straightforward calculations, the propagator evaluated at caustics
(\ref{K_qq0b}). The resulting expressions are in agreement with
those obtained by other authors (see, for example,
\cite{Lawande1,Lawande2,Rezende1,ChengChan}), though in our case
they have been attained within a different scheme, based essentially
on the previous obtention of a closed expression for the evolution
operator. Finally, a direct calculation shows that the propagator
$K(q,t;q_0,t_0)$, viewed as a function of $(q,t)$, formally
satisfies the evolution equation
$$
i\partial_t K =-\frac{1}{2}\partial_q^2
K+\frac{1}{2}q^2\kappa(t)K\,.
$$

\subsection{Transition amplitudes and vacuum
instability}\label{TransInstab}

\indent The exact expressions for the Green functions (\ref{K_qq0a})
and (\ref{K_qq0b}) can be used to \emph{exactly} compute both
transition amplitudes and probabilities. Here, we will restrict
ourselves to the class of normalized states $\phi_n^\omega$ defined
in $L^2(\mathbb{R},\mathrm{d}q)$ as
\begin{equation}\label{phin}
\phi^\omega_n(q):=\frac{\omega^{1/4}}{\sqrt{2^n
n!\sqrt{\pi}}}\exp\Big(-\omega q^2/2\Big)
H_n(\sqrt{\omega}q)\,,\,\,\, \omega>0\,,\,\,\, n\in\mathbb{N}_0\,,
\end{equation}
with $H_n(z)$ denoting the $n$th Hermite polynomial in the variable
$z\,$. Note that for any fixed value $\omega$ the set
$(\phi_n^\omega\,:\,n\in\mathbb{N}_0)$ defines the usual orthonormal
basis of $L^2(\mathbb{R})$ constituted by the eigenvectors of the
quantum Hamiltonian (\ref{quantHamiltonian}) corresponding to a TIHO
of constant frequency $\sqrt{\kappa(t)}=\omega$. Since the
$\phi_n^\omega$ states are complete, the corresponding transition
amplitudes and probabilities for other states are readily
obtainable. By using the generating function for Hermite
polynomials,
$$
\exp\big(2\sqrt{\omega}qx-x^2\big)=\sum_{n=0}^\infty
H_n(\sqrt{\omega}q)\frac{x^n}{n!}\,,
$$
it is clear that
\begin{equation}\label{transition}
\langle\phi^{\omega_2}_{n_2}\,|\,U(t_2,t_1)\phi^{\omega_1}_{n_1}\rangle=\frac{1}{\pi}\left(\frac{n_1!n_2!\sqrt{\omega_1\omega_2}}
{2^{n_1+n_2+1} i
}\right)^{1/2}s^{-1/2}(t_2,t_1)\,[x_1^{n_1}x_2^{n_2}]\,I(x_1,x_2;\Lambda(t_1,t_2;\omega_1,\omega_2))\,,
\end{equation}
where $[x_1^{n_1}x_2^{n_2}]f(x_1,x_2)$ denotes the complex
coefficient appearing in the $x_1^{n_1}x_2^{n_2}$-term of the Taylor
expansion of the function $f$. Here, for any matrix
$\Lambda\in\mathrm{Mat}_{2\times2}(\mathbb{C})$, we define
\begin{eqnarray*}
I(x_1,x_2;\Lambda)&:=&\exp\big(-(x^2_1+x^2_2)\big)\int_{\mathbb{R}^2}\exp\left(-\frac{1}{2} \vec{q}\,^{t}\Lambda\,\vec{q}+2\vec{x}\,^{t}\,\mathrm{diag}(\sqrt{\omega_1},\sqrt{\omega_2}) \,  \vec{q}\right) \,\mathrm{d}^2\vec{q}\\
&=&\frac{2\pi}{\sqrt{\det\Lambda}}\exp\bigg(
\vec{x}\,^{t}\Big(2\mathrm{diag}(\sqrt{\omega_1},\sqrt{\omega_2})\Lambda^{-1}\mathrm{diag}(\sqrt{\omega_1},\sqrt{\omega_2})-\mathbb{I}\Big)
\vec{x} \bigg)\,,
\end{eqnarray*}
whenever $\mathrm{Re}(\Lambda)\geq 0\,$ and ${\det}\Lambda\ne0\,$.
In this formula, $\vec{x}$ denotes the column vector with first and
second components given by $x_1$ and $x_2$, respectively; we define
$\vec{q}$ similarly. In our case,
\begin{eqnarray*}
 \Lambda(t_1,t_2;\omega_1,\omega_2):=\left(
             \begin{array}{cc}
               \omega_1-i\displaystyle\frac{c(t_2,t_1)}{s(t_2,t_1)} &\displaystyle \frac{i}{s(t_2,t_1)} \\
              \displaystyle \frac{i}{s(t_2,t_1)} & \omega_2-i\displaystyle\frac{c(t_1,t_2)}{s(t_2,t_1)}\\
             \end{array}
           \right),
\end{eqnarray*}
with
$$\det\Lambda(t_1,t_2;\omega_1,\omega_2)=
\left(\omega_1\omega_2-\frac{\dot{c}(t_2,t_1)}{s(t_2,t_1)}\right)-i\left(\frac{\omega_1c(t_1,t_2)+\omega_2
c(t_2,t_1)}{s(t_2,t_1)}\right).
$$
Here, $\mathrm{Re}(\Lambda(t_0,t;\omega_1,\omega_2))\ge0$ and
$\det\Lambda(t_0,t;\omega_1,\omega_2)\neq0$ for all $(t_0,t)\in
I\times I$ and $\omega_1,\omega_2\in(0,+\infty)\,$. The Taylor
expansion of $I(x_1,x_2;\Lambda)$ can be efficiently computed by
applying the following lemma, that trivially follows from the
multinomial formula.

\begin{lm}
Let
$$B=B^t=\left(\begin{array}{cc}b_{11} & b_{12}\\ b_{12} & b_{22}\end{array}\right)\in \mathrm{Mat}_{2\times 2}(\mathbb{C})\,.
$$
Then, using the notation introduced above, we have
\begin{eqnarray*}
[x_1^{n_1}x_2^{n_2}]\exp\bigg(\vec{x}\,^t B
\vec{x}\bigg)=b_{11}^{(n_1-n_2)/2}(2b_{12})^{n_2} \sum_{m\in
\Delta(n_1,n_2)} \frac{(b_{11}b_{22})^m
(4b^2_{12})^{-m}}{m!(m+(n_1-n_2)/2)!(n_2-2m)!}\,,
\end{eqnarray*}
whenever $n_1$ and $n_2$ have the same parity, and vanishes
otherwise. Here, $\Delta(n_1,n_2):=\big(m\in \mathbb{N}_0\,:\,
\max\{0,(n_2-n_1)/2\}\leq m\leq \lfloor n_2/2\rfloor\,\big)\,$,
where $\lfloor x\rfloor$ denotes the largest integer less than or
equal to $x\in\mathbb{R}\,$. In particular, taking $n_1=0\,$, we get
\begin{equation}\label{b22}
[x_1^{0}x_2^{n_2}]\exp\bigg(\vec{x}\,^t B
\vec{x}\bigg)=\frac{b_{22}^{n_2/2}}{(n_2/2)!}\quad\mathrm{for}\quad
n_2\equiv 0 \,\,\,(\mathrm{mod}\, 2)\,,
\end{equation}
and vanishes if $n_2$ is an odd number.
\end{lm}
\noindent \textbf{Remarks.} Note that the TDHO quantum dynamics is
invariant under parity inversion $\mathbf{P}$ and the states
$\phi_n^\omega$ satisfy $\mathbf{P}\phi_n^\omega=
(-1)^n\phi^\omega_n$. Hence,
$\langle\phi^{\omega_2}_{n_2}\,|\,U(t_2,t_1)\phi^{\omega_1}_{n_1}\rangle=0$
if $n_1$ and $n_2$ have different parity.
\\
\linebreak \indent As a concrete example, in the case of a TIHO with
constant frequency $\omega=\omega_1=\omega_2$, we identify
$$
B=2\,\mathrm{diag}(\sqrt{\omega},\sqrt{\omega})\,
\Lambda^{-1}(t_1,t_2;\omega,\omega)\,
\mathrm{diag}(\sqrt{\omega},\sqrt{\omega})-\mathbb{I}
=\exp\big(-i\omega(t_2-t_1)\big) \left(\begin{array}{cc} 0 & 1\\ 1 &
0
\\\end{array}\right)
$$
and, hence,
$$
I(x_1,x_2;\Lambda(t_1,t_2;\omega,\omega))=\sum_{n=0}^\infty\frac{2^n}{n!}\exp\big(-i\omega
n (t_2-t_1)\big) x_1^nx_2^n\,.
$$
This is in perfect agreement with
$$
\langle\phi^\omega_{n_2}\,|\,U(t_2,t_1)\phi^\omega_{n_1}\rangle=\exp\big(-i\omega(n_1+1/2)(t_2-t_1)\big)\delta(n_1,n_2)\,,
$$
where $\delta(n_1,n_2)$ denotes the Kronecker delta. For arbitrary
time-dependent frequencies the formula (\ref{transition}), when
restricted to the same initial and final frequencies
$\omega_1=\omega_2$, coincides with the one given in
\cite{Landovitz} written in terms of associated Legendre functions.
\\
\indent We conclude this section with the analysis of the
instability of the vacuum state $\phi_0^\omega$ due to the
nonautonomous nature of the Hamiltonian (\ref{quantHamiltonian}).
This can be easily derived from the formulae (\ref{transition}) and
(\ref{b22}).

\begin{thr}\label{th_vacuum}
The quantum time evolution of the vacuum state $\phi_0^\omega$ is
generally given by a superposition of states
$U(t,t_0)\phi_0^\omega=\sum_{n\in\mathbb{N}_0}\langle\phi_{2n}^\omega\,|\,U(t,t_0)\phi_0^\omega\rangle\phi_{2n}^\omega\,,$
where the probability amplitudes
$\langle\phi_{2n}^\omega\,|\,U(t,t_0)\phi_0^\omega\rangle$ are given
by
\begin{equation}\label{2n0}
\langle
\phi_{2n}^\omega\,|\,U(t,t_0)\phi_0^\omega\rangle=\frac{\sqrt{(2n)!}}{2^n
n!}\big(2\omega
(\Lambda^{-1}(t_0,t;\omega,\omega))_{22}-1\big)^n\langle\phi_0^\omega\,|\,U(t,t_0)\phi_0^\omega\rangle\,,\,\,\,n\in\mathbb{N}\,,
\end{equation}
in terms of the the expectation value
$$\langle\phi_0^\omega\,|\,U(t,t_0)\phi_0^\omega\rangle=\sqrt{\frac{2\omega}{\det\Lambda(t_0,t;\omega,\omega)}}\exp(-i\pi/4)\,s^{-1/2}(t,t_0)\,,$$
with
$$(\Lambda^{-1}(t_0,t;\omega,\omega))_{22}=\frac{\omega s^2(t_2,t_1)-i
s(t_2,t_1)c(t_2,t_1)}{1+\omega^2s^2(t_2,t_1)-c(t_2,t_1)c(t_1,t_2)-i\omega
s(t_2,t_1)\big(c(t_2,t_1)+c(t_1,t_2)\big)}\,.$$
\end{thr}
\noindent \textbf{Remarks.} Consider the usual annihilation and
creation operators
\begin{equation}\label{aa*usual}
a_\omega:=\frac{1}{\sqrt{2}}\big(\sqrt{\omega}Q+iP/\sqrt{\omega}\big)\quad
\mathrm{and} \quad
a_\omega^*:=\frac{1}{\sqrt{2}}\big(\sqrt{\omega}Q-iP/\sqrt{\omega}\big)
\,,
\end{equation}
with $[a_\omega,a_\omega^*]=\mathbf{1}$ and
$[a_\omega,a_\omega]=0=[a_\omega^*,a_\omega^*]$, such that
$a_\omega^*\phi_n^\omega=\sqrt{n+1}\phi_{n+1}^\omega$ and
$a_\omega\phi_n^\omega=\sqrt{n}\phi_{n-1}^\omega$,
$\forall\,n\in\mathbb{N}$, with $a_\omega\phi_0^\omega=0\,$. The
evolution of these operators in the Heisenberg picture can be
obtained directly from Eq. (\ref{QHPH}),
\begin{eqnarray}\label{BogolTransf}
U^{-1}(t,t_0)\,a_\omega\,U(t,t_0)&=&A_\omega(t,t_0)a_\omega+B_\omega(t,t_0)a_\omega^{*}\,,\\
U^{-1}(t,t_0)\,a_\omega^*\,U(t,t_0)&=&\bar{B}_\omega(t,t_0)a_\omega+\bar{A}_\omega(t,t_0)a_\omega^{*}\,,\nonumber
\end{eqnarray}
where $A_\omega(t,t_0)$ and $B_\omega(t,t_0)$ are the Bogoliubov
coefficients
\begin{eqnarray}
A_\omega(t,t_0)&:=&\frac{1}{2}\Big(c(t,t_0)+\dot{s}(t,t_0)+i\big(\omega^{-1}\dot{c}(t,t_0)-\omega
s(t,t_0)\big)\Big)\,,\label{BogolCoeff1}\\
B_\omega(t,t_0)&:=&\frac{1}{2}\Big(c(t,t_0)-\dot{s}(t,t_0)+i\big(\omega^{-1}\dot{c}(t,t_0)+\omega
s(t,t_0)\big)\Big)\,,\label{BogolCoeff2}
\end{eqnarray}
satisfying $A_\omega(t,t_0)=\bar{A}_\omega(t_0,t)$,
$B_\omega(t,t_0)=-B_\omega(t_0,t)$, and
$|A_\omega(t,t_0)|^2-|B_\omega(t,t_0)|^2=1\,$, $\forall\,(t,t_0)\in
I\times I\,$. Note, in particular, that $A_\omega(t,t_0)$ never
vanishes. For example, for the TIHO of constant frequency $\omega>0$
we have $B_\omega(t,t_0)=0$ and
$A_\omega(t,t_0)=\exp(-i(t-t_0)\omega)$. A straightforward
calculation yields (see also \cite{Parker} and \cite{PilchWarner})
\begin{equation}\label{vacuumevol}
U(t,t_0)\phi_0^\omega
=\langle\phi_0^\omega\,|\,U(t,t_0)\phi_0^\omega\rangle\exp\left(-\frac{1}{2}\frac{B_\omega(t_0,t)}{A_\omega(t_0,t)}a_\omega^{*2}\right)\phi_0^\omega\,,
\end{equation}
This formula is in perfect agreement with the transitions
(\ref{2n0}). Indeed, it is straightforward to check that
$$2\omega(\Lambda^{-1}(t_0,t;\omega,\omega))_{22}-1=-B_\omega(t_0,t)/A_\omega(t_0,t)\,.$$
Since $\det\Lambda(t_0,t;\omega,\omega)=-2i\omega
s^{-1}(t,t_0)A_\omega(t_0,t)$, the expectation value
$\langle\phi_0^\omega\,|\,U(t,t_0)\phi_0^\omega\rangle$ can be
rewritten as
\begin{equation}\label{Up0U}
\langle\phi_0^\omega\,|\,U(t,t_0)\phi_0^\omega\rangle=\frac{1}{\sqrt{|A_\omega(t_0,t)|}}\exp(i\sigma(t,t_0))\,,
\end{equation}
where the phase $\sigma(t,t_0)\in C^1(I\times I)$ comes from a
careful calculation of the principal argument. For a TIHO with
constant frequency $\omega>0$, we have
$\sigma(t,t_0)=(t_0-t)\omega/2$ for all $t,t_0\in\mathbb{R}\,$.
Given an arbitrary squared frequency $\kappa(t)$, the phase
$\sigma(t,t_0)$ evaluated at times $t$ close to $t_0$ is simply
given by
\begin{equation}\label{sigma}
\sigma(t,t_0)=-\frac{1}{2}\arctan\left(\frac{\omega
s(t,t_0)-\omega^{-1}\dot{c}(t,t_0)}{c(t,t_0)+\dot{s}(t,t_0)}\right).
\end{equation}
The $\sigma$ phase can be conveniently canceled through a suitable
redefinition of the Hamiltonian (\ref{quantHamiltonian}) just in the
case when $\dot{\sigma}(t,t_0)$ is independent of $t_0\,$. In that
situation, by identifying $\vartheta(t)=\dot{\sigma}(t,t_0)$ in Eq.
(\ref{K1}), we have that the redefined evolution operator satisfies
$\langle\phi_0^\omega\,|\,U_1(t,t_0)\phi_0^\omega\rangle=1/\sqrt{|A_\omega(t_0,t)|}\,$.
In the TIHO case, we get $\vartheta(t)=-\omega/2$ (this amounts to
considering normal order). In general, it is not possible to proceed
in this way in all situations when dealing with arbitrary
time-dependent frequencies. In any case, the $\sigma$ phase is
irrelevant for the calculation of transition probabilities. In
particular, given $\Psi_1,\Psi_2\in\mathscr{H}$ with
$\Psi_1=F_1(a_\omega,a_\omega^*)\,\phi_0^\omega\,$, where $F_1$ is
some analytic function, we have
\begin{eqnarray*}
\mathrm{Prob}(\Psi_2,t_2\,|\,\Psi_1,t_1)&=&|\langle\Psi_2\,|\,U(t_2,t_1)\Psi_1\rangle|^2\\
&=&\frac{|\langle\Psi_2\,|\,F_1(a_{\omega H}(t_1,t_2),a_{\omega
H}^{*}(t_1,t_2))\exp\big(-B_\omega(t_1
,t_2)/(2A_\omega(t_1,t_2))a_\omega^{*2}\big)\phi_0^\omega\rangle|^2}{|A_\omega(t_1,t_2)|}\,,
\end{eqnarray*}
where the time dependence only appears through the Bogoliubov
coefficients (\ref{BogolCoeff1}) and (\ref{BogolCoeff2}). Finally,
it is important to point out that the transformations
(\ref{BogolTransf}) and the evolution of the vacuum state
(\ref{vacuumevol}) fully characterize the quantum time evolution of
the TDHO. By using these relations, we can easily compute the action
of $U(t,t_0)$ on any basic vector
$\phi_n^\omega=(1/\sqrt{n!})\,a_\omega^{*n}\phi_0^\omega\,$.

%-------------------------------------------------
%                  SECTION IV
%-------------------------------------------------

\section{Semiclassical states}\label{SemiclStates}

\indent In this section, we will look for states that behave
semiclassically under the dynamics defined by the quantum
Hamiltonian (\ref{quantHamiltonian}). We will base our study on the
concrete factorization of the evolution operator defined in
\emph{theorem \ref{ThrU}}. To achieve this goal, note that the
eigenvalue problem for the Lewis invariant (\ref{Lewis}) can be
exactly solved. Indeed, let us fix $t_0\in I$ and let
$(\phi_n\,:\,n\in\mathbb{N}_0)$ be the eigenstates (\ref{phin}) of
the auxiliary  Hamiltonian $H_0$ (\ref{H0}) corresponding to unit
frequency $\omega=1$. According to the relation (\ref{TIT-1}), the
initial states $\psi^{\rho}_n(t_0):=T_\rho^{-1}(t_0)\phi_n$,
$$\psi^{\rho}_n(t_0,q)=\left(\frac{1}{2^n n!\sqrt{\pi}\rho(t_0)}\right)^{1/2}\exp\left(\frac{i}{2}\left(\frac{\dot{\rho}(t_0)}{\rho(t_0)}
+\frac{i}{\rho^2(t_0)}\right)q^2\right)H_n(q/\rho(t_0))\in
L^2(\mathbb{R},\mathrm{d}q)\,,$$ labeled both  by $\rho$ and the
integers $n\in\mathbb{N}_0$, are eigenstates of $I_\rho(t_0)$ with
eigenvalues equal to $n+1/2$. Consider now the initial pure state
\begin{equation}\label{semiclass}
\Phi_{\rho}^{(z)}(t_0):=T_\rho^{-1}(t_0)\Phi^{(z)}
=e^{-|z|^{2}/2}\sum_{n=0}^{\infty}\frac{z^n}{\sqrt{n!}}\psi^\rho_n(t_0)\,,\,\,\,\,z\in\mathbb{C}\,,
\end{equation}
with
$\Phi^{(z)}:=e^{-|z|^2/2}\sum_{n=0}^{\infty}\frac{z^n}{\sqrt{n!}}\phi_n$
being the well-known coherent states for the Hamiltonian $H_0\,$.
Let us take the annihilation and creation operators $a$ and $a^*$
for unit frequency $\omega=1$ defined in Eq. (\ref{aa*usual}). The
superposition (\ref{semiclass}) is a normalized eigenvector of the
(time-dependent) annihilation operator
\begin{equation}\label{arho1D}
a_\rho(t_0):=T_\rho^{-1}(t_0)\,a\,T_\rho(t_0)
=\frac{1}{\sqrt{2}}\big(Q/\rho(t_0)+i(\rho(t_0)P-\dot{\rho}(t_0)Q)\big)\,,
\end{equation}
in the sense that
$a_\rho(t_0)\Phi^{(z)}_\rho(t_0)=z\Phi^{(z)}_\rho(t_0)\,$. This
operator, together with the associated creation operator
$$a^*_\rho(t_0):=T_\rho^{-1}(t_0)\,a^*\,T_\rho(t_0)
=\frac{1}{\sqrt{2}}\big(Q/\rho(t_0)-i(\rho(t_0)P-\dot{\rho}(t_0)Q)\big)\,,$$
satisfies the Heisenberg algebra,
$[a_\rho(t_0),a^*_\rho(t_0)]=\mathbf{1}$ and
$[a_\rho(t_0),a_\rho(t_0)]=0=[a^*_\rho(t_0),a^*_\rho(t_0)]$, for
each initial value of the time parameter $t_0\,$. In particular, the
Lewis invariant (\ref{Lewis}) may be expressed in terms of these
operators as
$I_\rho(t_0)=a^*_\rho(t_0)a_\rho(t_0)+(1/2)\mathbf{1}\,$. Through
unitary time evolution, we get
\begin{equation}\label{Phit}
\Phi_\rho^{(z)}(t,t_0):=U(t,t_0)\Phi_{\rho}^{(z)}(t_0)
=\exp\left(-\frac{i}{2}\int_{t_0}^{t}\frac{\mathrm{d}\tau}{\rho^{2}(\tau)}\right)\Phi_{\rho}^{(z_\rho(t,t_0))}(t)\,,
\end{equation}
where we have denoted
$$
z_\rho(t,t_0):=\exp\left(-i\int_{t_0}^{t}\frac{\mathrm{d}\tau}{\rho^2(\tau)}\right)\,z\,,\quad
z\in\mathbb{C}\,.$$ We want to remark that the time-dependent phase
appearing in Eq. (\ref{Phit}) is necessary for these states to
verify the Schr\"{o}dinger equation. In our case, they coincide with
those defined in Eq. (4.6) of reference \cite{HartleyRay}. We
conclude that the family of states (\ref{semiclass}) is closed under
the dynamics. Moreover, the following theorem can be used to justify
that these states can be considered as semiclassical under certain
assumptions.

\begin{thr}\label{th_semiclass}
Let $z=x+iy\in \mathbb{C}$ and $t_0\in I$. The position and momentum
expectation values in the state
$\Phi_\rho^{(z)}(t,t_0)=U(t,t_0)\Phi_{\rho}^{(z)}(t_0)$ satisfy
\begin{eqnarray*}
q_H(t,t_0)&=&\big\langle\Phi_{\rho}^{(z)}(t,t_0)\,\big|\,Q\,\Phi_{\rho}^{(z)}(t,t_0)\big\rangle
=\sqrt{2}\rho(t)\,\mathrm{Re}\big(z_\rho(t,t_0)\big)\,,\\
p_H(t,t_0)&=&\big\langle\Phi_{\rho}^{(z)}(t,t_0)\,\big|\,P\Phi_{\rho}^{(z)}(t,t_0)\big\rangle
=\sqrt{2}\,\mathrm{Re}\big(\big(\dot{\rho}(t)-i/\rho(t)\big)z_\rho(t,t_0)\big)\,,
\end{eqnarray*}
where $(q_H,p_H)$ is the classical solution (\ref{Hamilton_class})
determined by the Cauchy data $(q,p)=\big(\sqrt{2}\rho(t_0)
x,\sqrt{2}( \dot{\rho}(t_0)x+y/\rho(t_0))\big)$ at time $t_0\,$.
Moreover, the mean square deviations of the position and momentum
operators with respect to the evolved state
$\Phi_\rho^{(z)}(t,t_0)$,
\begin{eqnarray}\label{varianceQP}
\Delta_{\Phi_{\rho}^{(z)}(t,t_0)}Q=\frac{1}{\sqrt{2}}\,\rho(t)\,,\quad
\Delta_{\Phi_{\rho}^{(z)}(t,t_0)}P=\frac{1}{\sqrt{2}}\,\big|\dot{\rho}(t)-i\rho^{-1}(t)\big|\,,
\end{eqnarray}
are independent of both $t_0$ and the Cauchy data defined by $z\,$.
\end{thr}

\begin{rem} \emph{Given any observable $\mathcal{O}$,  its uncertainty in the
state $\Psi\in\mathscr{D}_\mathcal{O}$ is defined as $\Delta_\Psi
\mathcal{O}:=\big(\langle\Psi\,|\,\mathcal{O}^2\Psi\rangle-\langle\Psi\,|\,\mathcal{O}\Psi\rangle^2\big)^{1/2}$.
Note that, in general, the elements of the family of states under
consideration are neither standard coherent states nor squeezed
states. For instance, for the free particle (\ref{c&sfree}) one can
choose $\rho(t)=\sqrt{1+(t-t_0)^2}$ and, hence,
$\Delta_{\Phi_{\rho}^{(z)}(t,t_0)}Q\sim t/\sqrt{2}$ for large values
of $t\,$; similar results occur for other elections of $\rho\,$.
Nevertheless, it is obvious that we will obtain good semiclassical
states for a system whenever the solution $\rho$ to the auxiliary EP
equation (\ref{EP_eq}) has a suitable behavior, for instance, if
$\rho$ is periodic in time or is simply a bounded function. We will
analyze some clarifying examples in this respect.}
\end{rem}

\begin{figure}
  \includegraphics[width=16cm]{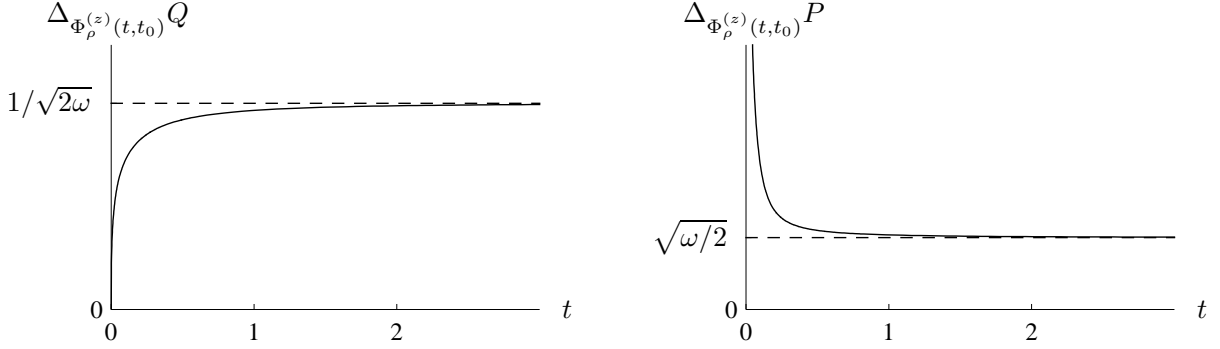}\\
  \caption{Variances of the position and momentum operators for the 3-torus Gowdy-type oscillator. Here,
  $\rho(t)=\sqrt{\pi t\big(J_0^2(\omega t)+Y_0^2(\omega t)\big)/2}$. The $\Phi^{(z)}_\rho(t,t_0)$ are states of minimum uncertainty
  for times $t$ far from the singularity at $t=0$\,.}\label{GowdyT3}
\end{figure}

\begin{figure}
  \includegraphics[width=16cm]{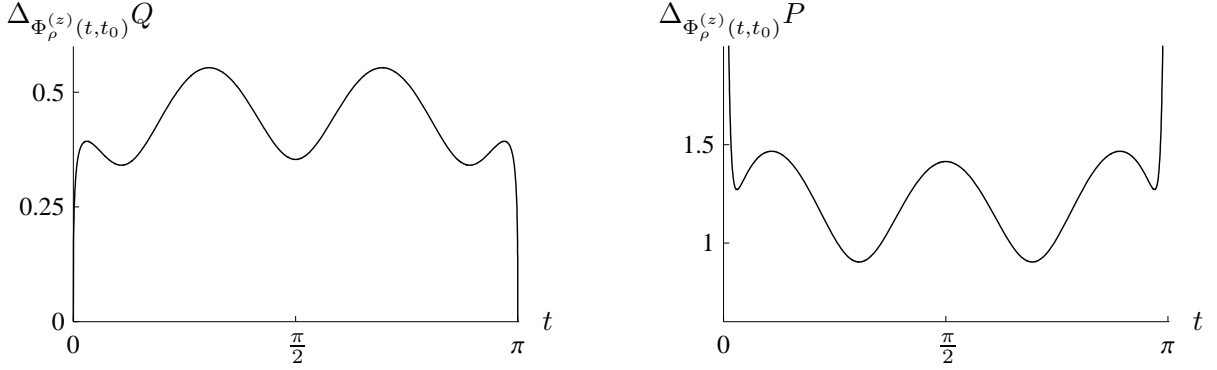}\\
  \caption{Variances of the position and momentum operators for the 3-handle and 3-sphere Gowdy-type oscillators.
  Here, we take the solution $\rho(t)=\sqrt{\sin t}\big(\mathscr{P}^2_{(\omega^\prime-1)/2}(\cos t)+\mathscr{Q}^2_{(\omega^\prime-1)/2}(\cos t)\big)^{1/2}$ to the auxiliar
  Ermakov-Pinney equation. In particular, these graphics correspond to $\omega^{\prime}=5\,$.
}\label{GowdyS1S2&S3}
\end{figure}

\noindent \textbf{Example 1} (Vertically driven pendulum)\textbf{.}
Consider the vertically driven pendulum \cite{Jose&Saletan}, i.e.,
the motion of a physical pendulum whose supporting point oscillates
in the vertical direction. In the small angles regime, it is
described by the Mathieu equation in its canonical form
\cite{Mathieu}
$$\ddot{u}(t)+\kappa(t;a,b)u(t)=0\,,\quad \kappa(t;a,b):=a-2b\cos(2t)\,,\quad a,b\in\mathbb{R}\,.$$
The general solution to this equation is a real linear combination
of the so-called Mathieu cosine and sine functions
\cite{McL,Abramow}, denoted respectively as $Ce(t;a,b)$ and
$Se(t;a,b)$. Given a nonzero $b$ value, it is a well-known fact that
the Mathieu cosine and sine functions are periodic in the time
parameter $t$ only for certain (countable number of) values of the
$a$ parameter, called \emph{characteristic values}. The procedure to
calculate these characteristic values for even or odd Mathieu
functions with \emph{characteristic exponent}\footnote{All Mathieu
functions have the form $\exp(irt)F(t)$, where $r$ is the
characteristic exponent and the function $F(t)$ has period $2\pi$.}
$r\in\mathbb{Z}$ and parameter $b$ can be efficiently implemented in
a computer. In this case, solutions to the EP equation (\ref{EP_eq})
inherit the periodic behavior from the Mathieu solutions, in such a
way that one obtains well-behaved semiclassical states for which the
average position and momentum follow the classical trajectories,
whereas the corresponding uncertainties vary periodically in time.
Note that, for small values of the $b$ parameter, we have
$Ce(t;a,b)\sim\cos(\sqrt{a}t)$ and $Se(t;a,b)\sim\sin(\sqrt{a}t)$,
and the system closely approximates the TIHO with squared frequency
given by the $a$ parameter.
\\
\linebreak \noindent \textbf{Example 2} ($\mathbb{T}^{3}$ Gowdy-like
oscillator)\textbf{.} Consider the TDHO equation
$$\ddot{u}(t)+\kappa(t;\omega)u(t)=0\,,\quad \kappa(t;\omega):=\omega^2+\frac{1}{4t^2}\,,\,\,\,\omega\in\mathbb{R}\,,\,\,\,t\in(0,+\infty)\,.$$
This equation is satisfied for each mode of the scalar fields
encoding the information about the gravitational local degrees of
freedom of the so-called $\mathbb{T}^3$ Gowdy models, which are
symmetry reductions of general relativity with cosmological
interpretation that admit an exact --i.e., nonperturbative--
quantization (see next section). In terms of the zero Bessel
functions of first and second kind \cite{Abramow}, denoted $J_0$ and
$Y_0$ respectively, the $c$ and $s$ solutions introduced in
\emph{section \ref{TDHOproperties}} are given by
\begin{eqnarray}
c(t,t_0)\!\!&=&\!\!\frac{\pi}{4}\Big(\sqrt{\frac{t}{t_0}}Y_0(\omega
t_0)\!-\!2\omega \sqrt{t_0t}Y_1(\omega t_0)\Big) J_0(\omega t) -
\frac{\pi}{4}\Big(\sqrt{\frac{t}{t_0}}J_0(\omega t_0)-2\omega
\sqrt{t_0t}J_1(\omega t_0)\Big) Y_0(\omega t),\nonumber\\
s(t,t_0)\!\!&=&\!\!-\frac{\pi}{2}\sqrt{t_0t}Y_0(\omega
t_0)J_0(\omega t)+\frac{\pi}{2}\sqrt{t_0t}J_0(\omega t_0)Y_0(\omega
t)\,.\label{c&s3torus}
\end{eqnarray}
Note that the squared frequency is a sum of a positive constant
$\omega^2$ plus a decreasing function of time, so that the system
approaches a time-independent oscillator as $t$ tends to infinity.
In Fig. \ref{GowdyT3}, we show states $\Phi_{\rho}^{(z)}(t,t_0)$
that behave as coherent states for large values of the time
parameter. The classical equation of motion has a singularity at
$t=0$ which translates into the vanishing of the uncertainty of the
position operator --and, hence, into the divergence of the variance
for the conjugate momentum-- at that instant of time.
\\
\indent There are other interesting effects due to the classical
singularity. Let us consider again the study of transition
amplitudes developed in \emph{subsection \ref{TransInstab}} and take
$\omega_1=\omega_2=\omega$. We proceed to analyze the behavior of
the (unique) state $\Psi(t_2,t_1)$ that evolves to the vacuum state
$\phi_0^\omega$ at time $t_2$ when used as Cauchy data in
$t_1<t_2\,$, i.e.,
$$
U(t_2,t_1)\Psi(t_2,t_1)=\phi^\omega_0\Leftrightarrow
\Psi(t_2,t_1)=U(t_1,t_2)\phi_0^\omega\,.
$$
The transition amplitudes $\langle
\phi^\omega_{2n}\,|\,\Psi(t_2,t_1)\rangle = \langle
\phi^\omega_{2n}\,|\,U(t_1,t_2)\phi_0^\omega\rangle$,
$n\in\mathbb{N}_0$, can be computed by using Eq. (\ref{2n0}). We
recognize two regions of interest in the time domain,
\begin{eqnarray*}
T_{0+}:=\{(t_1,t_2)\,\,|\,\,0<t_1\ll\omega^{-1}\ll t_2\}
\quad\mathrm{and}\quad T_{++}:=\{(t_1,t_2)\,\,|\,\,\omega^{-1}\ll
t_1< t_2\}\,.
\end{eqnarray*}
In $T_{++}$, the asymptotic behavior of the Bessel functions for
large values of the time parameter \cite{Abramow} leads the system
to behave as a TIHO of constant frequency $\omega$, with
$\Psi(t_2,t_1)\sim \phi_0^\omega\,$. On the other hand, in the
region $T_{0+}\,$, the proximity of $t_1$ to the classical
singularity manifests itself in the fact that the wave function
takes the form $\Psi(t_2,t_1)\sim 0\,$. Note that this behavior is
in conflict with the unitary evolution of the system, which implies
$\|\Psi(t_2,t_1)\|=1\,$.
\\
\linebreak \noindent \textbf{Example 3}
($\mathbb{S}^1\times\mathbb{S}^2$ and $\mathbb{S}^3$ Gowdy-like
oscillators)\textbf{.} Gowdy models admit spatial topologies
different from the 3-torus one, concretely the 3-handle
$\mathbb{S}^1\times\mathbb{S}^2$ and the 3-sphere $\mathbb{S}^3$. As
expected for closed universes, these systems present both initial
and final singularities. For this reason, they become useful test
beds to discuss the exact quantization of cyclic universes. Here,
the modes satisfy equations of motion of the form
$$\ddot{u}(t)+\kappa(t;\omega)u(t)=0\,,\quad \kappa(t;\omega):=\omega^2+\frac{1}{4}(1+\csc^2 t)\,,\,\,\,\omega\in\mathbb{R}\,,\,\,\,t\in(0,\pi)\,.$$
In this case, in terms of first and second class Legendre functions
\cite{Abramow} denoted respectively as $\mathscr{P}_x$ and
$\mathscr{Q}_x$, $x\in\mathbb{R}\,$, we have
\begin{eqnarray}\label{c&sremaintopol}
c(t,t_0)&=&\frac{1}{2}\sqrt{\sin t/\sin
t_0}\mathscr{Q}_{(\omega^\prime-1)/2}(\cos
t)\big((1+\omega^{\prime})\mathscr{P}_{(1+\omega^\prime)/2}(\cos
t_0)
-\omega^\prime\mathscr{P}_{(\omega^\prime-1)/2}(\cos t_0)\big)\nonumber\\
&-&\frac{1}{2}\sqrt{\sin t/\sin
t_0}\mathscr{P}_{(\omega^\prime-1)/2}(\cos
t)\big((1+\omega^{\prime})\mathscr{Q}_{(1+\omega^\prime)/2}(\cos
t_0)-\omega^\prime\mathscr{Q}_{(\omega^\prime-1)/2}(\cos
t_0)\big)\,,\nonumber\\
s(t,t_0)&=&\sqrt{\sin t\sin
t_0}\big(\mathscr{Q}_{(\omega^\prime-1)/2}(\cos
t_0)\mathscr{P}_{(\omega^\prime-1)/2}(\cos
t)-\mathscr{P}_{(\omega^\prime-1)/2}(\cos
t_0)\mathscr{Q}_{(\omega^\prime-1)/2}(\cos t)\big)\,,\nonumber\\
\end{eqnarray}
where $\omega^\prime:=\sqrt{1+4\omega^2}\,$. In Fig.
\ref{GowdyS1S2&S3}, we show the behavior of states
$\Phi_{\rho}^{(z)}(t,t_0)$ for which the uncertainties of the
position and momentum operators have an oscillatory behavior far
enough from the singularities occurring at $t=0$ and $t=\pi$.
Although $\rho$ does not vary periodically, the function remains
bounded and, thus, the $\Phi_{\rho}^{(z)}(t,t_0)$ states can be used
to perform a semiclassical study of these models. Finally, one may
proceed as in the 3-torus case in order to analyze the way the
classical singularities affect the quantum behavior of the systems,
obtaining similar results.

%-------------------------------------------------
%                   SECTION V
%-------------------------------------------------

\section{Extension to field theories}\label{FieldTheory}

\subsection{General framework}

\indent In this section, we will extend the previous study to linear
dynamical systems with infinite degrees of freedom, focusing our
attention on the study of scalar fields evolving in fixed
backgrounds such as the Minkowskian space or the so-called (linearly
polarized) Gowdy metrics \cite{Gowdy}, that correspond to symmetry
reductions of general relativity describing cosmological models with
initial and final singularities.\footnote{The isometry group of the
Gowdy models is $U(1)\times U(1)$ and the spatial slices are
restricted to have the topology of a 3-torus $\mathbb{T}^3$, a
3-handle $\mathbb{S}^1\times\mathbb{S}^2$, a the 3-sphere
$\mathbb{S}^3$, or the lens spaces $L(p,q)$ (that can be studied by
imposing discrete symmetries on the 3-sphere case). The exact
quantization of the linearly polarized Gowdy models in the vacuum or
coupled to massless scalar fields has been profusely analyzed (see
\cite{Corichi} and \cite{Barbero3} and references therein).} Our
results, however, will have a wide range of applicability, being
possible to easily extend them to other field theories with similar
structure. The construction of the appropriate $C^{*}$-algebra of
quantum observables can be obtained by making a simple comparison
with the one-dimensional case discussed in \emph{section
\ref{Unitaryoperator}}. Consider the canonical phase space
$(\Gamma,\bm{\omega})$ consisting of the infinite-dimensional
$\mathbb{R}$-vector space $\Gamma$ of smooth Cauchy data endowed
with the natural (weakly) symplectic structure $\bm{\omega}$. Taking
the linearity of $\Gamma$ into account, each element
$\lambda\in\Gamma$ is identified with the functional
$F_{\lambda}:\Gamma\rightarrow\mathbb{R}$ such that, for any other
$\lambda^\prime\in\Gamma$,
$F_{\lambda}(\lambda^\prime):=\bm{\omega}(\lambda,\lambda^\prime)\,$.
The abstract quantum algebra of observables is then given by the
Weyl $C^{*}$-algebra on $\Gamma$, $\mathscr{W}(\Gamma)$, generated
by the elements $W(\lambda)=\exp(iF_{\lambda})$, $\lambda\in\Gamma$,
that formally satisfy the relations (\ref{Weyl}). The GNS
construction \cite{Wald} establishes that, given any state
$\omega_0:\mathscr{W}(\Gamma)\rightarrow\mathbb{C}$ on the algebra
--that is, a normalized positive linear functional--, there exist a
Hilbert space
$(\mathscr{H}_{0},\langle\cdot|\cdot\rangle_{\mathscr{H}_0})$, a
representation
$\pi_{0}:\mathscr{W}(\Gamma)\rightarrow\mathscr{B}(\mathscr{H}_{0})$
from the Weyl algebra to the collection of bounded linear operators
on $\mathscr{H}_{0}$, and a cyclic vector
$\Psi_{0}\in\mathscr{H}_{0}$ such that
$$\omega_{0}(A)=\langle\Psi_{0}\,|\,\pi_{0}(A)\Psi_{0}\rangle_{\mathscr{H}_0}\,,\,\,\,\forall\,A\in\mathscr{W}(\Gamma)\,.$$
Moreover, the triplet $(\mathscr{H}_{0},\pi_{0},\Psi_{0})$
satisfying these properties is unique up to unitary equivalence. Von
Neumann's uniqueness theorem, however, cannot be generalized to
field theories, and states on the Weyl algebra generally yield
nonequivalent cyclic representations. In order to pick out a
preferred representation of the canonical commutation relations one
can impose additional criterions, such as the well-known Poincar\'e
invariance of the Fock representation for scalar fields in
Minkowskian space. Regarding the Gowdy models, one imposes the
invariance under an extra $U(1)$ symmetry generated by a residual
global constraint for the 3-torus case \cite{UniquenessMena}, or the
invariance under the symmetry group $SO(3)$ of the Klein-Gordon
equations of motion for the 3-handle and the 3-sphere cases
\cite{Barbero3}.
\\
\indent Every linear symplectic transformation
$\mathcal{T}\in\mathrm{SP}(\Gamma)$, for which
$\bm{\omega}(\mathcal{T}\lambda_1,\mathcal{T}\lambda_2)
=\bm{\omega}(\lambda_1,\lambda_2)\,$,
$\lambda_1,\lambda_2\in\Gamma\,$, defines a unique $*$-automorphism
$\alpha_{\mathcal{T}}\in\mathrm{Aut}(\mathscr{W}(\Gamma))$ such that
$(\alpha_{\mathcal{T}}\circ W)(\lambda):=W(\mathcal{T}\lambda)\,$.
This is the case, in particular, of the symplectic transformations
that characterize the classical dynamics of the system. Given a
concrete Hilbert space representation $(\mathscr{H}_0,\pi_0,\Psi_0)$
of the Weyl $C^*$-algebra, the symplectic transformation
$\mathcal{T}\in\mathrm{SP}(\Gamma)$ is said to be \emph{unitarily
implementable} on the cyclic representation space $\mathscr{H}_0$ if
$\pi_0$ and $\pi_0\circ\alpha_\mathcal{T}$ are unitarily equivalent,
i.e., there exists a unitary operator
$U_{\mathcal{T}}:\mathscr{H}_0\rightarrow\mathscr{H}_0$ such that
\begin{equation}\label{unitarityT}
U_{\mathcal{T}}^{-1}\pi(W(\lambda))U_{\mathcal{T}}=\pi((\alpha_{\mathcal{T}}\circ
W)(\lambda))\,,\,\,\,\, \forall\,
W(\lambda)\in\mathscr{W}(\Gamma)\,.
\end{equation}
A common feature of the quantization of infinite-dimensional linear
symplectic dynamical systems is, precisely, the impossibility of
defining the unitary quantum counterpart of \emph{all} linear
symplectic transformations on the phase space \cite{Shale}. This is
the case of the time evolution of the Gowdy models when these
systems are written in terms of the dynamical variables that
naturally appear after performing their Hamiltonian formalisms.
Note, however, that the lack of a unitary operator implementing the
quantum time evolution conflicts with the axiomatic structure of
quantum theory itself. In case of not rejecting the models for this
reason, one must carefully analyze the viability of a suitable
probabilistic interpretation for them, as discussed in \cite{Torre}.
Nevertheless, it is possible to overcome this problem by performing
some suitable time-dependent redefinitions of the basic fields
\cite{Corichi,Barbero3}. In what follows, we will always refer to
these rescaled fields.

\subsection{Unitary quantum time evolution}{\label{Unitary quantum time evolution}}

\indent The canonical phase space description of the classical
systems under consideration consists now of an infinite-dimensional
nonautonomous Hamiltonian system
$(I\times\Gamma,\mathrm{d}t,\bm{\omega},H(t))$. Here,
$\Gamma:=\mathscr{C}\times \mathscr{C}$ is the space of Cauchy data,
where $\mathscr{C}$ denotes the Fr\'echet space of rapidly
decreasing real sequences $\mathbf{x}:=(x_{\ell}\,:\,\ell\in
\mathbb{X})$, with $\ell$ running over some countable set
$\mathbb{X}$ \cite{Simon,Glimm,Gelfand}. This space is endowed with
the natural symplectic structure
$$\bm{\omega}((\mathbf{q}_1,\mathbf{p}_1),(\mathbf{q}_2,\mathbf{p}_2))
:=\sum_{\ell\in\mathbb{X}}\big(p_{1\ell}q_{2\ell}-p_{2\ell}q_{1\ell}\big)\,,\quad
\forall\,(\mathbf{q}_1,\mathbf{p}_1),
(\mathbf{q}_2,\mathbf{p}_2)\in\Gamma\,.
$$
The time-dependent Hamiltonian $H(t):\Gamma\rightarrow\mathbb{R}$ is
a quadratic form on $\Gamma$ that can be diagonalized as a sum of
TDHO Hamiltonians of the type (\ref{Hclass}). Explicitly,
\begin{equation}\label{Hamilton_fields}
H(t,\mathbf{q},\mathbf{p}):=\frac{1}{2}\sum_{\ell\in
\mathbb{X}}\big(p_\ell^2+\kappa_\ell(t)q^2_\ell\big)\,,\,\,\,\,t\in
I=(t_{-},t_{+})\subseteq\mathbb{R}\,,
\end{equation}
where the time-dependent squared frequencies $\kappa_\ell(t)\in
C^0(I)$, $\ell\in \mathbb{X}$, must satisfy
$\bm{\kappa}(t)=(\kappa_\ell(t)\,:\, \ell\in \mathbb{X})\in
\overline{\mathscr{C}}$, for all $t\in I$, with
$\overline{\mathscr{C}}$ denoting the vector space of slowly
increasing real sequences, i.e., the topological dual of
$\mathscr{C}$. The time evolution is implemented by symplectic
transformations of the type (\ref{Hamilton_class})
\begin{eqnarray*}
\left(\begin{array}{c}
                 q_{\ell H}(t,t_0) \\
                 p_{\ell H}(t,t_0) \\
               \end{array}
             \right)=\mathcal{T}^{(\ell)}_{(t,t_0)}\cdot\left(
               \begin{array}{c}
                 q_\ell \\
                 p_\ell \\
               \end{array}
             \right),\quad \mathcal{T}^{(\ell)}_{(t,t_0)}:=\left(
        \begin{array}{cc}
          c_\ell(t,t_0) & s_\ell(t,t_0) \\
       \dot{c}_\ell(t,t_0)& \dot{s}_\ell(t,t_0) \\
        \end{array}
      \right),\quad \ell\in\mathbb{X}\,,
\end{eqnarray*}
where, for each $\ell\in\mathbb{X}$, $c_\ell$ and $s_\ell$ are the
solutions to the TDHO equation of squared frequency $\kappa_\ell(t)$
introduced in \emph{section \ref{TDHOproperties}}.
\\
\indent For the Minkowskian quantum field theory generalized to a
spacetime $\mathbb{R}\times\mathbb{T}^3$ with closed spatial
sections we have $\mathbb{X}=\mathbb{Z}\setminus\{0\}$ and
$\kappa_\ell(t)=|\ell|^2$, for all $t\in\mathbb{R}$. For the 3-torus
Gowdy models \cite{Corichi} we take
$\mathbb{X}=\mathbb{Z}\setminus\{0\}$ and
$\kappa_\ell(t)=|\ell|^2+1/(4t^2)\,$, with $t\in(0,+\infty)$. For
the remaining topologies admitted by the Gowdy cosmologies, the
3-handle and the 3-sphere \cite{Barbero3}, $\mathbb{X}=\mathbb{N}$
and $\kappa_\ell(t)=\ell(\ell+1)+(1+\csc^2 t)/4\,$, with
$t\in(0,\pi)\,$. For simplicity, we will not consider in these cases
the quantization of the zero mode $\ell=0$. It can be represented in
terms of standard position and momentum operators with dense domains
in $L^2(\mathbb{R})\,$. The $c_\ell$ and $s_\ell$ functions for the
Minkowskian and Gowdy 3-torus cases are respectively given by
(\ref{c&sMinkowski}) and (\ref{c&s3torus}) substituting
$\omega=|\ell|\,$; for the 3-handle and 3-sphere topologies, they
are given by (\ref{c&sremaintopol}) identifying
$\omega=\sqrt{\ell(\ell+1)}\,$. The classical singularities of the
Gowdy models will persist in their quantum formulations.
\\
\indent In order to exactly quantize the infinite-dimensional
systems under consideration, we introduce Schr\"{o}dinger
representations \cite{CCQSchr,TorreSchr,MenaSchr,MyselfSchr}, where
state vectors act as functionals
$\Psi:\overline{\mathscr{C}}\rightarrow\mathbb{C}$ belonging to
certain Hilbert spaces
$\mathscr{H}_{\bm{\alpha}}=L^2(\overline{\mathscr{C}},\sigma(\mathrm{Cyl}(\mathscr{C})),\mathrm{d}\mu_{\bm{\alpha}})\,$,
and define suitable decompositions of the position and momentum
operators in terms of modes. There are subtleties associated with
the infinite-dimensionality of the classical configuration space
$\mathscr{C}$ that affect the definition of the Hilbert space. On
one hand, it is not possible to define nontrivial Lebesgue-type
translation invariant measures $\mu_{\bm{\alpha}}$, but rather
probability ones \cite{Gelfand,Yamasaki}. On the other hand, the set
over which the measure space is built --the so-called quantum
configuration space-- must be given by some suitable distributional
extension of $\mathscr{C}$. In this case, it suffices to consider
the dual $\overline{\mathscr{C}}$. The reason to proceed in this way
is that the measure is not supported on $\mathscr{C}$, i.e., the
classical configuration space has zero measure with respect to
$\mu_{\bm{\alpha}}$. Given a nonzero complex sequence
$\bm{\alpha}=(\alpha_{\ell}\,:\,\ell\in \mathbb{X})$, the Gaussian
measure $\mu_{\bm{\alpha}}$ is defined on the cylinder sets
$\sigma$-algebra $\sigma(\mathrm{Cyl}(\overline{\mathscr{C}}))$ on
$\overline{\mathscr{C}}$. This is the smallest $\sigma$-algebra with
respect to which the functionals
$\mathbf{q}\in\overline{\mathscr{C}}\mapsto \langle
\mathbf{q},\mathbf{x}\rangle:=\sum_{\ell\in\mathbb{X}} q_\ell
x_\ell$ are measurable for each $\mathbf{x}\in \mathscr{C}$.
Explicitly, for each finite $n$-tuple $(\ell_1,\dots,\ell_n)\in
\mathbb{X}^n$, such that $\ell_i<\ell_{i+1}$ ($i=1,\dots,n-1$), let
us consider the projections
$p_{\ell_1\cdots\ell_n}:\overline{\mathscr{C}}\rightarrow
\mathbb{R}^{n}$, $\mathbf{q}\mapsto
p_{\ell_1\cdots\ell_n}(\mathbf{q})=(q_{\ell_1},\dots,q_{\ell_n})\,$.
Then, $\mu_{\bm{\alpha}}$ is defined by its action on cylinder sets
belonging to $\sigma$-algebras of the form
$\mathcal{B}_{\ell_1\cdots
\ell_n}(\overline{\mathscr{C}})=p^{-1}_{\ell_1\cdots\ell_n}(\mathcal{B}(\mathbb{R}^n))\subset
\sigma(\mathrm{Cyl}(\overline{\mathscr{C}}))$, where
$\mathcal{B}(\mathbb{R}^n)$ is the Borel $\sigma$-algebra of subsets
of $\mathbb{R}^n$, i.e.,
\begin{equation*}
\left.\mathrm{d}\mu_{\bm{\alpha}}\right|_{\mathcal{B}_{\ell_1\cdots
\ell_n}}=\prod_{i=1}^n\frac{1}{\sqrt{2\pi}\,|\alpha_{\ell_i}|}\exp\left(-\frac{q_{\ell_i}^2}{2|\alpha_{\ell_i}|^2}\right)
\mathrm{d}q_{\ell_i}\,.
\end{equation*}
The sequence $\bm{\alpha}$ defines a positive continuous
nondegenerate bilinear form
$C_{\bm{\alpha}}:\mathscr{C}\times\mathscr{C}\rightarrow\mathbb{R}$
through the formula
$C_{\bm{\alpha}}(\mathbf{x},\mathbf{y}):=\sum_{\ell\in\mathbb{X}}|\alpha_\ell|^2x_\ell
y_\ell\,$, $\mathbf{x},\mathbf{y}\in\mathscr{C}\,$.
$C_{\bm{\alpha}}$ is called the \emph{covariance} operator in this
context \cite{Glimm}.
\\
\indent We consider now certain class of measurable functions that
we will use in the following. Let $\mathscr{X}\subset \mathscr{C}$
be a finite-dimensional subspace of the classical configuration
space $\mathscr{C}$. A cylinder function $\Psi$, based on
$\mathscr{X}$, is a function of the form
$$
\Psi(\mathbf{q})=F(\langle
\mathbf{q},\mathbf{x}_1\rangle,\ldots,\langle
\mathbf{q},\mathbf{x}_n\rangle)
$$
for a finite set $\mathbf{x}_1,\ldots,\mathbf{x}_n \in \mathscr{X}$,
where $F:\mathbb{R}^n\rightarrow \mathbb{C}$ is a smooth function.
They are called cylinder because they depend on $\mathbf{q}\in
\overline{\mathscr{C}}$ through the pairings $\langle
\mathbf{q},\mathbf{x}_i\rangle$ defined by a finite number of
``probes'' $\mathbf{x}_i\in \mathscr{C}$. The Gaussian  measure
$\mu_{\bm{\alpha}}$ can be used to endow the linear space spanned by
this type of functions with an inner product. Explicitly, in the
particular case
$\mathscr{X}=\mathrm{span}\{\mathbf{e}_{\ell_1},\ldots,\mathbf{e}_{\ell_n}\}$,
where $\mathbf{e}_i\in \mathscr{C}$ denotes a sequence whose only
nonzero component is the $i$th one, every cylinder function $\Psi$
on $\mathscr{X}$ can be written in the form
$\Psi(\mathbf{q})=F(q_{\ell_1},\ldots,q_{\ell_n})$ and the scalar
product is given by
$$
\langle \Psi_1\,|\,\Psi_2 \rangle=\int_{\mathbb{R}^n}
\bar{F}_1(q_{\ell_1},\ldots,q_{\ell_n}
)F_2(q_{\ell_1},\ldots,q_{\ell_n} )
\prod_{i=1}^n\frac{1}{\sqrt{2\pi}\,|\alpha_{\ell_i}|}\exp\left(-\frac{q_{\ell_i}^2}{2|\alpha_{\ell_i}|^2}\right)
\mathrm{d}q_{\ell_i}\,.
$$
Square integrable cylinder functions on $\mathscr{X}$ span a
$\mathbb{C}$-vector space that we denote as
$\mathrm{Cyl}_{\mathscr{X}}$. The class of all cylinder functions is
denoted by $\mathrm{Cyl}:=\cup_{\mathscr{X}}
\mathrm{Cyl}_{\mathscr{X}}$. The inner products defined on each
$\mathrm{Cyl}_{\mathscr{X}}$ can be extended to $\mathrm{Cyl}$ in
the natural way and the Cauchy completion of $\mathrm{Cyl}$ with
respect to this inner product is the Hilbert space
$\mathscr{H}_{\bm{\alpha}}$ (see \cite{Yamasaki} for more details).
\\
\indent The configuration observables will act as multiplication
operators, whereas the canonically conjugate momenta will differ
from the usual ones in terms of derivatives by the appearance of
multiplicative terms which are necessary to ensure their
self-adjointness; specifically, for $\Psi\in\mathrm{Cyl}$,
\begin{equation}\label{Q&Pell}
(Q_\ell\Psi)(\mathbf{q})=q_\ell\Psi(\mathbf{q})\,,\quad
(P_\ell\Psi)(\mathbf{q})=-i\frac{\partial\Psi}{\partial
q_\ell}(\mathbf{q})+\frac{\bar{\beta}_\ell}{\bar{\alpha}_\ell}
q_\ell\Psi(\mathbf{q})\,.
\end{equation}
The complex sequences
$\bm{\alpha}$ and $\bm{\beta}$ must satisfy
\begin{equation}\label{alpha&beta}
\alpha_\ell\bar{\beta}_\ell-\beta_\ell\bar{\alpha}_\ell=i\,,\,\,\,
\forall\,\ell\in \mathbb{X}\,,
\end{equation}
by virtue of the CCR,
$[Q_\ell,P_{\ell^\prime}]=i\delta(\ell,\ell^\prime)\mathbf{1}$ and
$[Q_\ell,Q_{\ell^\prime}]=0=[P_\ell,P_{\ell^\prime}]$. These
conditions imply $|\alpha_\ell||\beta_\ell|\ge1/2$ for all
$\ell\in\mathbb{X}\,$. The functional form of $\bm{\alpha}$ and
$\bm{\beta}$ for the Minkowskian case and Gowdy models will be
discussed later in the context of the unitary implementability of
their dynamics.
\\
\indent According to condition (\ref{unitarityT}), if the quantum
dynamics is unitarily implementable there exists a (biparametric)
family of unitary operators
$U(t,t_0):\mathscr{H}_{\bm{\alpha}}\rightarrow
\mathscr{H}_{\bm{\alpha}}$, depending on $(t,t_0)\in I\times I$,
such that
\begin{eqnarray}
&&U^{-1}(t,t_0)\,Q_\ell\,U(t,t_0)=c_\ell(t,t_0)Q_\ell+s_\ell(t,t_0)P_\ell\,,\label{UQ}\\
&&U^{-1}(t,t_0)\,P_\ell\,U(t,t_0)=\dot{c}_\ell(t,t_0)Q_\ell+\dot{s}_\ell(t,t_0)P_\ell\,.\label{UP}
\end{eqnarray}
The above relations characterize $U(t,t_0)$ univocally up to phase.
They can be rewritten in terms of annihilation and creation
operators $a_\ell$ and $a^*_\ell\,$, with
$[a_\ell,a^*_{\ell^{\prime}}]=\delta(\ell,\ell^\prime)\mathbf{1}\,$
and $[a_\ell,a_{\ell^\prime}]=0=[a^*_\ell,a^*_{\ell^\prime}]\,$,
such that
\begin{eqnarray}
Q_\ell=\alpha_\ell a_\ell+\bar{\alpha}_\ell a^*_\ell\,,\,\,
P_\ell=\beta_\ell a_\ell+\bar{\beta}_\ell a^*_\ell\Leftrightarrow
a_\ell=-i\bar{\beta}_\ell Q_\ell+i\bar{\alpha}_\ell P_\ell\,,\,\,
a^*_\ell=i\beta_\ell Q_\ell-i\alpha_\ell P_\ell\,.\label{aa*}
\end{eqnarray}
Relations (\ref{UQ}) and (\ref{UP}) are then equivalent to
\begin{eqnarray}
U^{-1}(t,t_0)\,a_\ell\,U(t,t_0)&=&A_\ell(t,t_0)a_\ell +
B_\ell(t,t_0)a_\ell^*\,,\label{U-1aUfields}\\
U^{-1}(t,t_0)\,a_\ell^*\,U(t,t_0)&=&\bar{B}_\ell(t,t_0)a_\ell +
\bar{A}_\ell(t,t_0)a_\ell^*\,,\nonumber
\end{eqnarray}
where the Bogoliubov coefficients $A_\ell(t,t_0)$ and
$B_\ell(t,t_0)$ are given by
\begin{eqnarray}
A_\ell(t,t_0)&:=&i\Big(\dot{s}_\ell(t,t_0)\bar{\alpha}_\ell\beta_\ell
-c_\ell(t,t_0)\bar{\beta}_\ell\alpha_\ell
+\dot{c}_\ell(t,t_0)|\alpha_\ell|^2-s_\ell(t,t_0)|\beta_\ell|^2\Big)\,,\label{Aell}\\
B_\ell(t,t_0)&:=&
i\Big(\big(\dot{s}_\ell(t,t_0)-c_\ell(t,t_0)\big)\bar{\alpha}_\ell\bar{\beta}_\ell
+\dot{c}_\ell(t,t_0)\bar{\alpha}_\ell^2-s_\ell(t,t_0)\bar{\beta}_\ell^2\Big)\,,\label{Bell}
\end{eqnarray}
satisfying $|A_\ell(t,t_0)|^2-|B_\ell(t,t_0)|^2=1$, for all
$\ell\in\mathbb{X}\,$. In particular, $A_\ell(t,t_0)\neq0$ for all
$(t,t_0)\in I\times I\,$. According to the theory of unitary
implementation of symplectic transformations \cite{Shale}, the
unitary time evolution operator $U(t,t_0)$ exists if and only if
\begin{equation}\label{Unitcond}
\mathbf{B}(t,t_0)=(B_\ell(t,t_0)\,:\,\ell\in\mathbb{X})\in\ell^{2}(\mathbb{C})\,\Leftrightarrow\,
\sum_{\ell\in\mathbb{X}}|B_\ell(t,t_0)|^2<+\infty\,,\, \forall\,
(t,t_0)\in I\times I\,.
\end{equation}
For quantum free fields in Minkowskian spacetime the usual choice
\begin{equation}\label{f&gMinkowski}
\alpha_\ell=\frac{1}{\sqrt{2|\ell|}}\quad\mathrm{and}\quad
\beta_\ell=-i\sqrt{\frac{|\ell|}{2}}
\end{equation}
implies $\mathbf{B}(t,t_0)=\mathbf{0}$ and, as is well known, the
time evolution is unitarily implementable for inertial observers.
The uniqueness of this representation can be proved by using the
same techniques employed in the standard
$\mathbb{R}\times\mathbb{R}^3$ Minkowskian spacetime under the
condition of Poincar\'e invariance. Concerning the Gowdy models, it
is straightforward to show that the unitarity of the quantum time
evolution is guaranteed for $\bm{\alpha}$ and $\bm{\beta}$ sequences
with the asymptotic expansions (see \cite{Corichi,Barbero1,Barbero3}
for more details)
\begin{equation*}
\alpha_\ell=\frac{1}{\sqrt{2|\ell|}}\exp(i\gamma_\ell)+O\big(|\ell|^{-3/2}\big)\,,\quad
\beta_\ell=-i\sqrt{\frac{|\ell|}{2}}\exp(i\gamma_\ell)+O\big(|\ell|^{-1/2}\big)\,,
\end{equation*}
where $\bm{\gamma}$ is an arbitrary real-valued sequence. Moreover,
all ($U(1)$- or $SO(3)$-invariant) representations for which the
dynamics is well-defined and unitary are unitarily equivalent. In
what follows, we will assume the use of the particular choice of
$\bm{\alpha}$ and $\bm{\beta}$ coincident with Eq.
(\ref{f&gMinkowski}). As a counterexample to the previous cases with
unitary evolution, consider a system of infinitely many harmonic
oscillators with imaginary frequency of the type (\ref{c&stachy}).
In this case, it is possible to show the nonexistence of sequences
$\bm{\alpha}$ and $\bm{\beta}$ satisfying (\ref{alpha&beta}) such
that the dynamics is unitarily implemented. The `wrong sign' of the
squared frequency is also responsable for the failure of the
unitarity of the time evolution in more complicated systems, such as
minimally coupled massless scalar fields evolving in de Sitter
spacetime \cite{deSitter}.
\\
\indent Note that the annihilation operators are given by the
derivatives
$$
a_\ell=\bar{\alpha}_\ell\frac{\partial}{\partial q_\ell}
$$
and, hence, the vacuum state $\Psi_0\in\mathscr{H}_{\bm{\alpha}}$
satisfying $a_\ell \Psi_0=0$ for all $\ell\in \mathbb{X}$ is given
by the unit constant functional $\Psi_0(\mathbf{q})=1$ up to
multiplicative phase. The states with finite number of particles,
which are obtained as the image of any polynomial in the creation
operators acting on the vacuum state, define a common, invariant,
dense domain of analytic vectors for the configuration and momentum
operators (\ref{Q&Pell}), so that their essential self-adjointness
is guaranteed and, hence, the existence of unique self-adjoint
extensions (see Nelson's analytic vector theorem in
\cite{Reed&Simon}).
\\
\indent From the Bogoliubov transformations (\ref{U-1aUfields}), and
proceeding as in the one-dimensional case (\ref{vacuumevol}), one
easily computes the evolution of the vacuum state,
\begin{equation}\label{vacuumevolfield}
U(t,t_0)\Psi_0=\langle\Psi_0\,|\,U(t,t_0)\Psi_0\rangle\exp\left(-\frac{1}{2}\sum_{\ell\in
\mathbb{X}}\frac{B_\ell(t_0,t)}{A_\ell(t_0,t)}a_\ell^{*2}\right)\Psi_0\,,
\end{equation}
with
\begin{equation}\label{expectedfield}
|\langle\Psi_0\,|\,U(t,t_0)\Psi_0\rangle|=\prod_{\ell\in\mathbb{X}}\frac{1}{\sqrt{|A_\ell(t_0,t)|}}\,.
\end{equation}
Due to the unitary implementability of the dynamics, the square
summability of the sequence
$(B_\ell(t_0,t)/A_\ell(t_0,t)\,:\,\ell\in\mathbb{X})$ and the
convergence of $\sum_{\ell\in\mathbb{X}}\log|A_\ell(t_0,t)|$ are
guaranteed and, hence, the action of $U(t,t_0)$ is well defined over
states with finite number of particles. The phase of the expectation
value (\ref{expectedfield}), though being irrelevant to answer most
of the physical questions, can be explicitly calculated once a
quantum Hamiltonian has been fixed. The Hamiltonian verifies Eq.
(\ref{Uequa}) and coincides with the operator directly promoted from
the classical expression (\ref{Hamilton_fields}) modulo an arbitrary
$t$-dependent real term proportional to the identity which encodes
the choice of $U(t,t_0)$, i.e.,
\begin{equation*}
H(t)=\frac{1}{2}\sum_{\ell\in\mathbb{X}}\big(P_\ell^2+\kappa_\ell(t)Q_\ell^2+2\vartheta_\ell(t)\mathbf{1}\big)\,,
\end{equation*}
where the sequence $\vartheta_\ell(t)\in C^{0}(I)$,
$\ell\in\mathbb{X}\,$, is usually employed to avoid the appearance
of infinite phases. Analogously to the one-dimensional case, when
the dynamics is unitarily implementable, we define the time
evolution propagator through the relation
\begin{equation*}
\big(U(t,t_0)\Psi\big)(\mathbf{q})=\int_{\overline{\mathscr{C}}}K_{\bm{\alpha}\bm{\beta}}\big(\mathbf{q},t;\mathbf{q}_0,t_0\big)
\Psi(\mathbf{q}_0)\,\mathrm{d}\mu_{\bm{\alpha}}(\mathbf{q}_0)\,,
\end{equation*}
where a straightforward calculation formally provides
\begin{equation}\label{K_fields}
K_{\bm{\alpha}\bm{\beta}}\big(\mathbf{q},t;\mathbf{q}_0,t_0\big)
=\prod_{\ell\in\mathbb{X}}\sqrt{2\pi}|\alpha_\ell|\exp\left(\frac{i}{2}\left(\frac{\beta_\ell}{\alpha_\ell}q_{0\ell}^2
-\frac{\bar{\beta}_\ell}{\bar{\alpha}_\ell} q_{\ell}^2\right)\right)
K_\ell\big(q_{\ell},t;q_{0\ell},t_0\big)\exp\left(-i\int_{t_0}^{t}\mathrm{d}\tau\,\vartheta_\ell(\tau)\right),
\end{equation}
with $K_\ell$ denoting the propagator (\ref{K_qq0a}) associated with
the one-dimensional oscillator of squared frequency
$\kappa_\ell(t)$. The reader may wish to compare this expression
with Eq. (\ref{Kalphabeta}) corresponding to a single oscillator.
This formula coincides with the one obtained in \cite{Rezende2} when
restricted to generic finite-dimensional and time-dependent linear
Hamiltonian systems. For the sequences (\ref{f&gMinkowski}), Eq.
(\ref{K_fields}) provides the expectation value
$$\langle\Psi_0\,|\,U(t,t_0)\Psi_0\rangle=\prod_{\ell\in\mathbb{X}}\frac{1}{\sqrt{|A_\ell(t_0,t)|}}\exp\left(i\Big(\sigma_\ell(t,t_0)-\int_{t_0}^{t}\mathrm{d}\tau\,\vartheta_\ell(\tau)\Big)\right),$$
with
$$\sigma_\ell(t,t_0)=-\frac{1}{2}\arctan\left(\frac{|\ell|
s_\ell(t,t_0)-|\ell|^{-1}\dot{c}_\ell(t,t_0)}{c_\ell(t,t_0)+\dot{s}_\ell(t,t_0)}\right)
$$
for times $t$ close to $t_0$. For the Minkowskian free fields, as in
the one-dimensional case, the phases $\sigma_\ell(t,t_0)$ can be
exactly canceled for all $\ell\in\mathbb{Z}\setminus\{0\}$ just by
defining a normal ordered quantum Hamiltonian, which amounts to
choosing $\vartheta_\ell(t)=-|\ell|/2$. In general, however, it is
not possible to eliminate them. For the Gowdy cosmologies, attending
to the fact that the $c_\ell$ and $s_\ell$ functions tend to those
corresponding to the Minkowskian case for large values of $|\ell|$,
it is easy to check that normal ordering allows only the cancelation
of the phases at high frequencies, with
$\vartheta_\ell(t)\sim-|\ell|/2$ as $|\ell|\rightarrow+\infty\,$.
\\
\indent Since we are dealing with systems of infinite number of
uncoupled oscillators, one would expect that the analysis developed
in \emph{section \ref{Unitaryoperator}} for a single oscillator
would allow us to factorize the evolution operator in the form
\begin{equation}\label{UFock}
U(t,t_0)=T_{\bm{\rho}}^{-1}(t)R_{\bm{\rho}}(t,t_0)T_{\bm{\rho}}(t_0)\,,
\end{equation}
where, given a sequence
$\bm{\rho}(t)=(\rho_\ell(t)\,:\,\ell\in\mathbb{X})$ of solutions to
the auxiliary Ermakov-Pinney equations
$\ddot{\rho}_\ell+\kappa_\ell(t)\rho_\ell=1/\rho_\ell^3\,$, the
$T_{\bm{\rho}}(t)$ and $R_{\bm{\rho}}(t,t_0)$ operators are
univocally characterized up to phases by their action on
annihilation and creation operators,
\begin{eqnarray*}
&&T_{\bm{\rho}}^{-1}(t)\,a_\ell\,T_{\bm{\rho}}(t)=i\left(\beta_\ell\bar{\alpha}_\ell\rho_\ell(t)-\frac{\alpha_\ell\bar{\beta}_\ell}{\rho_\ell(t)}
-|\alpha_\ell|^2
\dot{\rho}_\ell(t)\right)a_\ell+i\left(\bar{\alpha}_\ell\bar{\beta}_\ell\Big(\rho_\ell(t)-\frac{1}{\rho_\ell(t)}\Big)-\bar{\alpha}_\ell^2
\dot{\rho}_\ell(t)\right)a_\ell^*\,,\\
&&R_{\bm{\rho}}^{-1}(t,t_0)\,a_\ell\,R_{\bm{\rho}}(t,t_0)=
\left(\cos\left(\int_{t_0}^{t}\frac{\mathrm{d}\tau}{\rho_\ell^{2}(\tau)}\right)-i(|\alpha_\ell|^2+|\beta_\ell|^2)\sin\left(\int_{t_0}^{t}\frac{\mathrm{d}\tau}{\rho_\ell^{2}(\tau)}\right)\right)a_\ell\nonumber\\
&&\hspace*{3.6cm}-\,\,i(\bar{\alpha}_\ell^2+\bar{\beta}_\ell^2)\sin\left(\int_{t_0}^{t}\frac{\mathrm{d}\tau}{\rho_\ell^{2}(\tau)}\right)a_\ell^{*}\,,
\end{eqnarray*}
and similarly for $a_\ell^{*}\,$. Again, the resulting unitary
evolution operator should be independent of the particular choice of
$\bm{\rho}$ since, as we have shown above, the propagator does not
depend on it. However, even in the case of $U(t,t_0)$ being
well-defined as unitary operator, the factorization (\ref{UFock})
may be ill-defined. This is, in fact, the case for free fields
evolving in Minkowskian and Gowdy-type spacetimes, as we will prove
below. Obviously, this does not prevent us from defining other
well-defined factorizations for $U(t,t_0)$ different from
(\ref{UFock}). A particularly convenient choice is the one made in
\cite{Barbero1} for the 3-torus Gowdy model. Calculations developed
there can be translated essentially unchanged into the remaining
Gowdy spatial topologies and the Minkowskian case (see the
\emph{appendix \ref{appendix}}). Here, however, we are interested in
the original factorization due to its implications for the search of
semiclassical states. According to \cite{Shale}, the necessary and
sufficient condition for $T_{\bm{\rho}}(t)$ to be unitary for each
value of $t$ is given by
\begin{eqnarray}
\sum_{\ell\in\mathbb{X}}\big|\alpha_\ell\beta_\ell\big(\rho_\ell(t)-1/\rho_\ell(t)\big)-\alpha_\ell^2
\dot{\rho}_\ell(t)\big|^2<+\infty\,,\,\,\,\,\forall\,t\in
I\,.\label{uniT}
\end{eqnarray}
Similarly, it is straightforward to show that $R_{\bm{\rho}}(t,t_0)$
is unitarily implementable if and only if
\begin{eqnarray}
\sum_{\ell\in\mathbb{X}}\Big|(\alpha_\ell^2+\beta^2_\ell)
\sin\left(\int_{t_0}^t\frac{\mathrm{d}\tau}{\rho_{\ell}^2(\tau)}\right)\Big|^2<+\infty\,,\,\,\,\,\forall\,(t,t_0)\in
I\times I\,.\label{uniR}
\end{eqnarray}
\noindent Equations (\ref{f&gMinkowski}) for quantum free fields in
Minkowskian and Gowdy-type spacetimes lead us to conclude that
conditions (\ref{uniT}) and (\ref{uniR}) are not satisfied and,
hence, neither $T_{\bm{\rho}}(t)$ nor $R_{\bm{\rho}}(t,t_0)$  are
unitary in those systems. In the case of $R_{\bm{\rho}}(t,t_0)$,
this conclusion follows readily, irrespective of $\bm{\rho}(t)$. For
$T_{\bm{\rho}}(t)$, a necessary condition for (\ref{uniT}) to be
satisfied is given by
$$\sum_{\ell\in\mathbb{X}}|\rho_\ell(t)-1/\rho_\ell(t)|^2<+\infty\,\,\Leftrightarrow\,\,
\lim_{|\ell|\rightarrow+\infty}\rho_\ell(t)=1\,,\,\,\,\,\forall\,t\in
I\,,$$ where we have taken into account the fact that the real
sequence $(\rho_\ell(t)\,:\,\ell\in\mathbb{X})$ is positive and
bounded for all $t\,$. According to Eq. (\ref{s_rho}), this implies
$s_\ell(t,t_0)\sim\sin C(t,t_0)$ as $|\ell|\rightarrow+\infty$,
where $C(t,t_0)$ is a \emph{nonzero} function which depends on the
system and whose form we do not need to specify. This is in conflict
with the asymptotic behavior of $s_\ell(t,t_0)$ for the systems
under study, given by $s_\ell(t,t_0)\sim0$ as
$|\ell|\rightarrow+\infty$ for all $(t,t_0)\in I\times I\,$.

\subsection{Semiclassical states}

\indent The explicit expression of the quantum unitary evolution for
the single harmonic oscillator as a product of unitary operators
(see \emph{theorem \ref{ThrU}} in \emph{section
\ref{Unitaryoperator}}) turned out to be very useful to construct
semiclassical states for some relevant one-dimensional dynamical
systems. In particular, as stated in  (\ref{TIT-1}), the operator
$T_{\bm{\rho}}(t)$ transforms the Lewis invariant (\ref{Lewis}) into
the time-independent free Hamiltonian (\ref{H0}). However, as we
have shown in \textit{subsection \ref{Unitary quantum time
evolution}}, there are obstructions that arise when dealing with
systems of infinite oscillators --particularly, the possible
nonunitarity of $T_{\bm{\rho}}(t)$--, making the application of the
techniques developed in \emph{section \ref{SemiclStates}}
particularly difficult. In order to avoid these difficulties, we
will probe an alternative procedure to construct semiclassical
states that takes advantage of the unitary implementability of the
quantum time evolution. We start by constructing the analogs of the
minimal wave packets of the one-dimensional harmonic oscillator.
Given a square summable sequence $\mathbf{z}=(z_\ell\,:\,\ell\in
\mathbb{X})\in\ell^2(\mathbb{C})$, consider the state
\begin{equation}\label{Cstates}
\Phi^{(\mathbf{z})}:=\mathrm{e}^{-\|\mathbf{z}\|^{2}/2}\exp\Big(\sum_{\ell\in\mathbb{X}}
z_{\ell}{a}_{\ell}^{*}\Big)\Psi_0\in \mathscr{H}_{\bm{\alpha}}\,,
\end{equation}
where the vacuum state $\Psi_0$ corresponds in this context to
$\mathbf{z}=0$, and
$\|\mathbf{z}\|=\sum_{\ell\in\mathbb{X}}|z_{\ell}|^{2}\,$. Vectors
defined in this way appear as coherent superpositions of states with
arbitrary number of particles. We then introduce the annihilation
and creation operators in the Heisenberg picture corresponding to
evolution \emph{backwards} in time,
\begin{eqnarray*}
a_{\ell}(t_0,t)&:=&U(t,t_0)\,a_{\ell}\,U^{-1}(t,t_0)=\bar{A}_{\ell}(t,t_0)a_{\ell}-B_{\ell}(t,t_0)a_{\ell}^{*}\,,\\
a^*_{\ell}(t_0,t)&:=&U(t,t_0)\,a_{\ell}^{*}\,U^{-1}(t,t_0)=-\bar{B}_{\ell}(t,t_0)a_{\ell}+
A_{\ell}(t,t_0)a_{\ell}^{*}\,,
\end{eqnarray*}
satisfying the Heisenberg algebra for all $(t,t_0)\in I\times I\,$.
Here, $A_\ell(t,t_0)$ and $B_\ell(t,t_0)$ are the Bogoliubov
coefficients defined in (\ref{Aell}) and (\ref{Bell}), respectively.
We then evolve the states (\ref{Cstates}) in the Schr\"{o}dinger
picture, obtaining
\begin{eqnarray*}
\Phi^{(\mathbf{z})}(t,t_0):=U(t,t_0)\,\Phi^{(\mathbf{z})}&=&
\mathrm{e}^{-\|\mathbf{z}\|^{2}/2}U(t_0,t)\exp\Big(\sum_{\ell\in\mathbb{X}}z_{\ell}a_{\ell}^{*}\Big)\Psi_0\\
&=&\mathrm{e}^{-\|\mathbf{z}\|^{2}/2}\exp\Big(\sum_{\ell\in\mathbb{X}}z_{\ell}a_{\ell}^{*}(t_0,t)\Big)\Phi^{(\bm{0})}(t,t_0)\,,
\end{eqnarray*}
with
$a_\ell(t_0,t)\,\Phi^{(\mathbf{z})}(t,t_0)=z_\ell\Phi^{(\mathbf{z})}(t,t_0)$,
$\forall\,\ell\in\mathbb{X}\,$, and
$\Phi^{(\bm{0})}(t,t_0)=U(t,t_0)\,\Psi_0\,$. By definition, the
one-parameter family of states obtained in this way verifies the
Schr\"{o}dinger equation with initial condition
$\Phi^{(\mathbf{z})}$, and is closed under time evolution as well,
$U(t_2,t_1)\,\Phi^{(\mathbf{z})}(t_1,t_0)=\Phi^{(\mathbf{z})}(t_2,t_0)\,$.
The states $\Phi^{(\mathbf{z})}(t,t_0)$  satisfy the properties
stated in the following theorem.

\begin{thr}\label{th_semiclass_fields}
Let $\mathbf{z}=(z_\ell\,:\,\ell\in
\mathbb{X})\in\ell^2(\mathbb{C})$ and $t_0\in I$. The position and
momentum expectation values in the state
$\Phi^{(\mathbf{z})}(t,t_0)=U(t,t_0)\Phi^{(\mathbf{z})}$ satisfy
\begin{eqnarray*}
q_{H
\ell}(t,t_0)&=&\big\langle\Phi^{(\mathbf{z})}(t,t_0)\,\big|\,Q_\ell\,\Phi^{(\mathbf{z})}(t,t_0)\big\rangle
=2c_\ell(t,t_0) \mathrm{Re}( \alpha_\ell z_\ell) +2 s_\ell(t,t_0)\mathrm{Re} (\beta_\ell z_\ell)\,,\\
p_{H
\ell}(t,t_0)&=&\big\langle\Phi^{(\mathbf{z})}(t,t_0)\,\big|\,P_\ell\,\Phi^{(\mathbf{z})}(t,t_0)\big\rangle
=2\dot{c}_\ell(t,t_0) \mathrm{Re} (\alpha_\ell z_\ell) +
2\dot{s}_\ell(t,t_0) \mathrm{Re} (\beta_\ell z_\ell)\,,
\end{eqnarray*}
where $(q_{H \ell},p_{H \ell})$ is the classical solution
(\ref{Hamilton_class}) determined by the Cauchy data
$(q_\ell,p_\ell)=\big(2\mathrm{Re} (\alpha_\ell z_\ell),2\mathrm{Re}
(\beta_\ell z_\ell))$ at time $t_0$. Moreover, the mean square
deviations of the position and momentum operators with respect to
the evolved state $\Phi^{(\mathbf{z})}(t,t_0)$ satisfy
\begin{eqnarray*}
\Delta_{\Phi^{(\mathbf{z})}(t,t_0)}Q_{\ell}=\big|\alpha_\ell
A_\ell(t,t_0)+\bar{\alpha}_\ell\bar{B}_\ell(t,t_0)\big|\,,\quad
\Delta_{\Phi^{(\mathbf{z})}(t,t_0)}P_{\ell}=\big|\beta_\ell
A_\ell(t,t_0)+\bar{\beta}_\ell\bar{B}_\ell(t,t_0)\big|\,,
\end{eqnarray*}
irrespective of $\mathbf{z}$.
\end{thr}

\begin{figure}[t] \centering
\includegraphics[width=15.5cm]{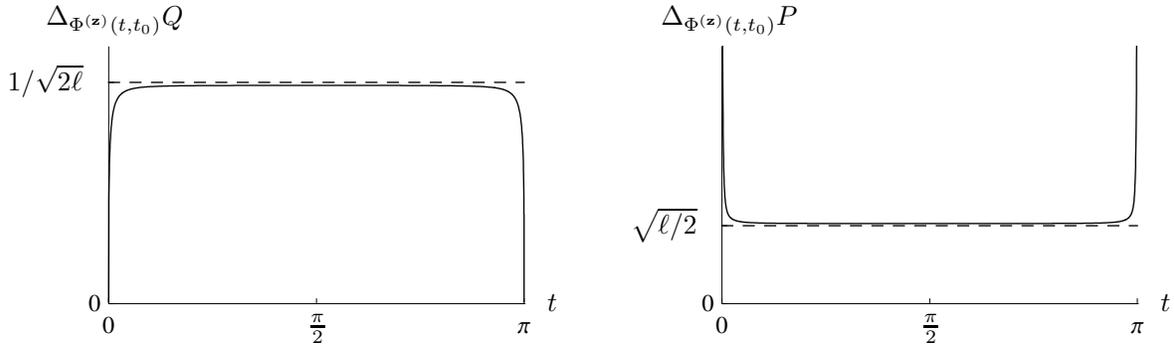}
\caption{Asymptotic behavior of the variances of the field and
momentum operators $Q_{\ell}$ and $P_{\ell}$ for the states
$\Phi^{(\mathbf{z})}(t,t_0)$, $t_0<t\in(0,\pi)$, at high frequencies
(limit $\ell\rightarrow+\infty$) for the 3-handle and 3-sphere Gowdy
models. Note the purely quantum behavior of these states as $t$
approaches the classical singularities. These graphics can be
considered as the limit of Fig. \ref{GowdyS1S2&S3} for large
frequencies. The asymptotic behavior for the 3-torus case is similar
to the one represented in Fig. \ref{GowdyT3}, identifying
$\omega=|\ell|$ for each mode.}\label{DeltaQ&Pfig}
\end{figure}

\begin{rem} \emph{Let us consider the Gowdy models. As a consequence
of the unitary implementability of the dynamics (\ref{Unitcond}), we
have $\mathbf{B}(t,t_0)\in\ell^2(\mathbb{C})$ and, hence,
$|A_\ell(t,t_0)|\sim1$ as $|\ell|\rightarrow+\infty$. We then obtain
(see Fig. \ref{DeltaQ&Pfig})
\begin{eqnarray}\label{asympt}
\Delta_{\Phi^{(\mathbf{z})}(t,t_0)}Q_{\ell}\sim|\alpha_\ell|=\frac{1}{\sqrt{2|\ell|}}\,,\quad
\Delta_{\Phi^{(\mathbf{z})}(t,t_0)}P_{\ell}\sim|\beta_\ell|=\sqrt{\frac{|\ell|}{2}}\quad\mathrm{when}\quad|\ell|\rightarrow+\infty\,.
\end{eqnarray}
For fixed values of $t_0$, these asymptotic behaviors converge
uniformly in $t$ for time intervals away from the classical
singularities. Note that these behaviors are the same that one would
have expected to obtain if it had been possible to suitably extend
the study developed in \emph{section \ref{SemiclStates}} to field
theories. We then conclude that the $\Phi^{(\mathbf{z})}(t,t_0)$
vectors are coherent states far enough from the singularities. For
the 3-torus model, the sequence $\mathbf{z}\in\ell^2(\mathbb{C})$ is
subject to satisfy a global constraint remaining on the system,
given by $\sum_{\ell\in\mathbb{Z}\setminus\{0\}}\ell|z_\ell|^2=0\,$.
In the Minkowskian case, expressions (\ref{asympt}) are valid for
all $\ell\in\mathbb{Z}\setminus\{0\}\,$.}
\end{rem}

%-------------------------------------------------
%                   SECTION VI
%-------------------------------------------------

\section{Conclusions}\label{Conclusions}

\indent In this paper, we have revised the unitary implementability
of the quantum dynamics of a time-dependent harmonic oscillator
(TDHO) and used the theory of invariants in order to define suitable
semiclassical states for some relevant systems, such as the
vertically driven pendulum or Gowdy-like oscillators. In particular,
we have analyzed some important issues related to the associated
Ermakov-Pinney equation, clarifying the need to introduce it as a
natural way to obtain an evolution operator valid for \emph{all}
values of the time parameter. We must emphasize again that other
elections different from the auxiliary Ermakov-Pinney equation may
be problematic because of the singular behavior of the resulting
evolution operator. We have shown that the Feynman propagator,
usually derived by making use of more complicated path-integration
techniques, can be obtained in a straightforward way within this
scheme. The resulting formula has then been applied to calculate
transition amplitudes, the instability of the vacuum state, and
semiclassical states. Most of the calculations regarding the quantum
evolution can be performed just by taking the classical dynamics
into account, except for the presence of a phase that depends on the
election of the quantum Hamiltonian. Nevertheless, this phase is
irrelevant to answer all relevant physical questions such as the
calculation of probability amplitudes or the evolution of quantum
observables. It is important to remark that this phase, in contrast
with the situation for the well-known TIHO system, cannot be
eliminated in all cases by considering normal ordered Hamiltonians.
\\
\indent Although we have concentrated our discussion on the quantum
TDHO, our results can be easily extended to another interesting
cases. For instance, our study is also applicable to the harmonic
oscillator driven by an external, nonstationary, perturbative force
characterized by a linear term in the position operator,
$$H(t)=\frac{1}{2}\big(P^2+\kappa(t)Q^2\big)+f(t)Q\,,\quad f\in C^1(I)\,.$$
Indeed, this Hamiltonian can be transformed into the Hamiltonian
(\ref{quantHamiltonian}) just by introducing an $f$-dependent
Glauber displacement operator \cite{MoyaCessa,Glauber2}.
\\
\indent Finally, we have extended the study of the unitary evolution
of a single quantum harmonic oscillator to systems of infinite
number of uncoupled oscillators with time-dependent frequencies,
concretely, to the quantum field theory in Minkowskian space and the
Gowdy cosmological models, providing a rigorous definition of the
propagator. Here, the impossibility of unitarily implementing some
symplectic transformations turns out to be an obstacle to generalize
the construction of semiclassical states through the eigenstates of
the Lewis invariant \cite{Lewis}. Nevertheless, we have shown that
the unitary implementability of the dynamics in appropriate
Schr\"{o}dinger representations allows us to define suitable
semiclassical states for these systems. In the case of the Gowdy
cosmologies, they can be used to probe the existence of large
quantum gravity effects in several ways. For instance, one may
construct suitable regularized operators to represent the (three- or
four-dimensional) metric of these models by using arguments similar
to those employed in the linearly polarized Einstein-Rosen waves
\cite{ERW1,ERW2,ERW3} and the Schmidt model \cite{Schmidt}.
Calculating the expectation values of these operators in the
coherent states, one may deduce the additional conditions (if any)
that the sequences $\mathbf{z}\in\ell^{2}(\mathbb{C})$ (see
\emph{theorem \ref{th_semiclass_fields}} in \emph{section
\ref{FieldTheory}}) should satisfy in order to admit an approximate
classical behavior. It is also important to analyze if the metric
quantum fluctuations are relevant for all states.
\\
\indent In addition, one may proceed as in \cite{4R} by
appropriately promoting the quadratic invariant
${^{\scriptscriptstyle{(4)}}}R_{abcd}{^{\scriptscriptstyle{(4)}}}R^{abcd}$
into a quantum mechanical operator. According to that reference, one
should be able to unambiguously fix the operator order by requiring
that the expectation values of this quantity in the coherent states
exactly reproduce the classical results far from the singularities.
In analogy with the results of \cite{4R}, even if the expectation
values in other states (such as linear combinations of coherent
states) give nonclassical results, it is expected that the classical
singularities persist in all cases. This physical consideration is
supported by the purely quantum behavior of the uncertainties of the
field and momentum operators in the coherent states at the classical
spacetime singularities.
\\
\indent The most natural way to extend the analysis developed in
this article consists in considering generic nonautonomous quadratic
Hamiltonians which contain the time-dependent harmonic oscillators
as particular cases. These systems can be analyzed from the
perspective of some recent works on this subject (see
\cite{Carinena1,Carinena2,Carinena3}) in which Lie systems in
quantum mechanics are studied from a geometrical point of view,
developing methods to obtain the time evolution operators associated
with time-dependent Schr\"{o}dinger equations of Lie-type. These
techniques may be successfully applied to infinite-dimensional
quadratic Hamiltonian systems by following a functional description
similar to the one performed in this article. In particular, the
different resulting factorizations for the time evolution operators
may be especially useful to define alternative families of
semiclassical states for these systems.

\begin{acknowledgments}

The authors are indebted to Fernando Barbero for his comments and
constant encouragement. D. G. Vergel also wishes to thank Prof. J.
Mour\~{a}o for his hospitality at the Instituto Superior T\'ecnico
(Lisbon), and acknowledges the support of the Spanish Research
Council (CSIC) through an I3P research assistantship. This work is
also supported by the Spanish MICINN research grant FIS2008-03221.

\end{acknowledgments}

%----------------------------------------------APPENDIX

\begin{appendix}

\section{Factorization of the evolution operator}\label{appendix}

\indent In this appendix, we briefly summarize the construction of a
well-defined factorization for the evolution operators of
Minkowskian free fields and Gowdy cosmologies. The reader is
referred to \emph{section \ref{FieldTheory}} to revise the notation
and definitions. A particularly useful way to proceed is to
factorize $U(t,t_0)$ as \cite{Barbero1}
$$U(t,t_0)=\mathcal{D}_{\bm{\rho}}(t,t_0)\mathcal{R}_{\bm{\rho}}(t,t_0)\mathcal{S}_{\bm{\rho}}(t,t_0)\,,$$
with
\begin{eqnarray*}
&&\mathcal{D}_{\bm{\rho}}(t,t_{0}):=D_{\bm{\rho}}^{-1}(t)D_{\bm{\rho}}(t_{0})\,,\\
&&\mathcal{S}_{\bm{\rho}}(t,t_{0}):=D_{\bm{\rho}}^{-1}(t_{0})
S_{\bm{\rho}}^{-1}(t)T_{\bm{\rho}}(t_{0})\,,\\
&&\mathcal{R}_{\bm{\rho}}(t,t_{0}):=T_{\bm{\rho}}^{-1}(t_{0})
R_{\bm{\rho}}(t,t_{0})T_{\bm{\rho}}(t_{0})\,,
\end{eqnarray*}
where $D_{\bm{\rho}}(t)$ and $S_{\bm{\rho}}(t)$ are displacement and
squeeze operators of the type defined in \emph{subsection
\ref{ConstructingU}}, in such a way that
\begin{eqnarray}
\mathcal{D}^{-1}_{\bm{\rho}}(t,t_{0})\,a_\ell\,\mathcal{D}_{\bm{\rho}}(t,t_{0})&=&\left(1+i|\alpha_\ell|^2\left(\frac{\dot{\rho}_\ell(t)}{\rho_\ell(t)}
-\frac{\dot{\rho}_\ell(t_0)}{\rho_\ell(t_0)}\right)\right)a_\ell+i\bar{\alpha}_\ell^2\left(\frac{\dot{\rho}_\ell(t)}{\rho_\ell(t)}-\frac{\dot{\rho}_\ell(t_0)}{\rho_\ell(t_0)}\right)a_\ell^*\,,\label{TransfD}
\\
\mathcal{S}^{-1}_{\bm{\rho}}(t,t_{0})\,a_\ell\,\mathcal{S}_{\bm{\rho}}(t,t_{0})&=&i\left(
\beta_\ell\bar{\alpha}_\ell\frac{\rho_\ell(t_0)}{\rho_\ell(t)}-\alpha_\ell\bar{\beta}_\ell\frac{\rho_\ell(t)}{\rho_\ell(t_0)}
+|\alpha_\ell|^2\frac{\dot{\rho}_\ell(t_0)}{\rho_\ell(t_0)}\left(\frac{\rho_\ell(t)}{\rho_\ell(t_0)}-\frac{\rho_\ell(t_0)}{\rho_\ell(t)}\right)\right)a_\ell
\nonumber\\
&+&i\bar{\alpha}_\ell\left(
\bar{\alpha}_\ell\frac{\dot{\rho}_\ell(t_0)}{\rho_\ell(t_0)}-\bar{\beta}_\ell\right)\left(\frac{\rho_\ell(t)}{\rho_\ell(t_0)}-\frac{\rho_\ell(t_0)}{\rho_\ell(t)}\right)a^*_\ell\,,\label{TransfS}
\\
\mathcal{R}^{-1}_{\bm{\rho}}(t,t_{0})\,a_\ell\,\mathcal{R}_{\bm{\rho}}(t,t_{0})&=&\Bigg(\cos\left(\int_{t_0}^{t}\frac{\mathrm{d}\tau}{\rho_\ell^{2}(\tau)}\right)
+i\Bigg((\alpha_\ell\bar{\beta}_\ell+\beta_\ell\bar{\alpha}_\ell)\dot{\rho}_\ell(t_0)\rho_\ell(t_0)\nonumber\\
&-&|\alpha_\ell|^2\left(\dot{\rho}^2_\ell(t_0)+\frac{1}{\rho_\ell^2(t_0)}\right)-|\beta_\ell|^2\rho_\ell^2(t_0)\Bigg)\sin\left(\int_{t_0}^{t}\frac{\mathrm{d}\tau}{\rho_\ell^{2}(\tau)}\right)\Bigg)\,a_\ell\nonumber\\
&+&i\Bigg(2\bar{\alpha}_\ell\bar{\beta}_\ell\dot{\rho}_\ell(t_0)\rho_\ell(t_0)-\bar{\alpha}_\ell^2\left(\dot{\rho}_\ell^2(t_0)+\frac{1}{\rho_\ell^2(t_0)}\right)
-\bar{\beta}_\ell^2\rho^2_\ell(t_0)\Bigg)\nonumber\\
&\times&
\sin\left(\int_{t_0}^{t}\frac{\mathrm{d}\tau}{\rho_\ell^{2}(\tau)}\right)a_\ell^{*}\,,\label{TransfR}
\end{eqnarray}
and similarly for $a_\ell^{*}\,$. Here, the solutions $\rho_\ell$ to
the EP equations are conveniently selected as follows. For the
Minkowskian free fields we choose $\rho_\ell(t)=1/\sqrt{|\ell|}\,$;
for the 3-torus Gowdy model, we take
\begin{equation}\label{rhoT3}
\rho_\ell(t)=\sqrt{\frac{\pi
t}{2}\big(J_0^2(|\ell|t)+Y_0^2(|\ell|t)\big)}\,,
\end{equation}
whereas for the 3-handle and the 3-sphere Gowdy models we choose
\begin{equation}\label{rhoS1xS2&S3}
\rho_\ell(t)=\sqrt{\frac{\sin t}{2}\left(\pi \mathscr{P}^2_\ell(\cos
t)+\frac{4}{\pi}\mathscr{Q}^2_\ell(\cos t)\right)}\,.
\end{equation}
Solutions (\ref{rhoT3}) and (\ref{rhoS1xS2&S3}) have the asymptotic
expansions
\begin{equation*}
\rho_\ell(t)=1/\sqrt{|\ell|}+O\big(|\ell|^{-3/2}\big)\,,\quad
\dot{\rho}_\ell(t)= C(t)/|\ell|^{5/2}+O\big(|\ell|^{-7/2}\big)\,,
\end{equation*}
as $|\ell|\rightarrow+\infty\,$. Here, $C(t)$ is a function of time
which depends on the spatial topology and whose form we do not need
to specify. It is then straightforward to check the unitary
implementability of the transformations (\ref{TransfD}),
(\ref{TransfS}), and (\ref{TransfR}) in the Hilbert space by
following the arguments employed in \emph{subsection \ref{Unitary
quantum time evolution}}.

\end{appendix}

\end{document}